\documentclass[%
 reprint,
reprint,onecolumn,11pt,
amsmath,amssymb,aps,
prb,
]{revtex4-1}

\usepackage{hyperref}
\usepackage{amsmath}
\usepackage{amssymb} 
\usepackage[pdftex]{graphicx}
\usepackage{braket} 
\normalfont \topmargin -0.5 cm
\begin{document}

\title{Dynamical suppression of
    fluctuations in an interacting nuclear spin bath of a self-assembled quantum dot using multiple pulse nuclear magnetic resonance}

\author{A. M. Waeber}
\email[]{andreas.waeber@wsi.tum.de}
\affiliation{Department of Physics and Astronomy, University of Sheffield, Sheffield S3 7RH, United Kingdom}
\affiliation{Walter Schottky Institut and Physik-Department, Technische Universit\"{a}t M\"{u}nchen, Am Coulombwall 4, 85748 Garching, Germany}
\author{M. Hopkinson}
\affiliation{Department of Electronic and Electrical Engineering, University of Sheffield, Sheffield S1 3JD, United Kingdom}
\author{M. S. Skolnick}
\affiliation{Department of Physics and Astronomy, University of Sheffield, Sheffield S3 7RH, United Kingdom}
\author{E. A. Chekhovich}
\email[]{e.chekhovich@sheffield.ac.uk}
\affiliation{Department of Physics and Astronomy, University of Sheffield, Sheffield S3 7RH, United Kingdom}

\date{\today}

\begin{abstract}
    Electron spin qubit coherence in quantum dots is ultimately
    limited by random nuclear spin bath fluctuations. Here we aim to
    eliminate this randomness by making spin bath evolution
    deterministic. We introduce spin bath control sequences, which
    systematically combine Hahn and solid echoes to suppress
    inhomogeneous broadening and nuclear-nuclear interactions.
    Experiments on self-assembled quantum dots show a five-fold
    increase in nuclear spin coherence. Numerical simulations show
    that these sequences can be used to suppress decoherence via
    qubit-qubit interaction in point defect and dopant spin systems.
\end{abstract}

\maketitle


The excellent spin-photon interface of confined charges in III-V
semiconductor quantum dots (QDs) has recently attracted a lot of
attention for potential applications in photon-mediated quantum
networks \cite{kimble2008,akopian2011,meyer2015}. The large
optical dipole moment of QDs makes ultrafast optical spin control
feasible and permits unrivalled entanglement generation rates
\cite{press2008,delteil2016,stockill2017}. On the other hand, the
coherence properties of the electron or hole spin qubit are
strongly affected by hyperfine interaction with the fluctuating
spin bath of the $\sim10^5$ constituent nuclei of the QD
\cite{khaetskii2002,merkulov2002}.

Recently, significant progress has been made in suppressing
hyperfine-induced qubit dephasing in strain-free QDs by applying
tailored dynamical decoupling protocols to the qubit spin
\cite{bluhm2010,malinowski2016}. However, there are limits to this
approach: in a strained QD, static quadrupolar fields lead to a
spread of the nuclear spin Larmor frequencies, reducing the
effectiveness of such a spectral filtering method
\cite{chekhovich2012,bechtold2015,stockill2016,waeber2016}.

In order to improve the properties of the qubit environment
directly, we explore the complementary pathway of controlling the
nuclear spin bath itself with pulsed nuclear magnetic resonance
(NMR) \cite{delange2012,perunicic2016}. Examples for controlling
spin-spin interactions are found in NMR spectroscopy where
sequences such as WAHUHA \cite{waugh1968} and MREV
\cite{rhim1973,mansfield1973} are used to average out dipolar
couplings selectively. However, these `solid echo' cycles do not
refocus inhomogeneous broadening caused by additional static or
time-dependent fields. On the other hand, dynamical decoupling
sequences consisting of a series of $\pi$-pulses can suppress this
inhomogeneous dephasing very effectively
\cite{cywinski2008,alvarez2010,ajoy2011}, but are unsuitable for
controlling dipolar coupling terms. Instead, such sequences even
increase dipolar dephasing through the parasitic effect of
instantaneous diffusion
\cite{klauder1962,raitsimring1974,tyryshkin2012}.

In this work, we introduce a set of pulse sequences which are
designed to combine the features of dynamical decoupling with
those of solid echoes. Under these \textit{combined Hahn and solid
    echo} (CHASE) multiple pulse sequences, we engineer the dynamics
of the interacting many-body nuclear spin bath in a single InGaAs
QD. While it is not possible to eliminate spin bath dynamics
completely, we show that random fluctuations can be transformed
into deterministic evolution, which can in principle be decoupled
from the qubit using standard control schemes
\cite{perunicic2016}. We test CHASE sequences experimentally in
optically detected nuclear magnetic resonance (ODNMR) measurements
and explore their applicability to systems with and without strong
inhomogeneous resonance broadening using first principle quantum
mechanical simulations. Our experimental results reveal an up to
fivefold increase in the CHASE nuclear spin coherence time of
$^{75}$As compared to the Hahn echo \cite{chekhovich2015}.
Furthermore, our simulations show that cyclic application of CHASE
sequences can suppress dephasing arbitrarily well in spin
ensembles with weak inhomogeneous broadening (e.g. strain-free
quantum dots, dilute donor spins or defect centres).

Before presenting experimental results, we describe how the
control sequences are designed. We start from an intuitive
approach previously used to extend nuclear spin lifetimes in
silicon and diamond \cite{ladd2005,maurer2012}: by combining solid
echo cycles with refocusing $\pi$-pulses, both inhomogeneous
dephasing and dipolar coupling can be suppressed. Here we
substantiate this approach by applying a rigorous average
Hamiltonian theory (AHT) \cite{haeberlen1968}, which is a form of
perturbation theory based on Magnus expansion. The evolution of a
spin-1/2 nuclear bath $I_i$ is analysed under a given pulse cycle
in an external magnetic field $B_\mathrm{z}$. We take into account
a dipolar coupling term $\mathcal{H}_\mathrm{d}^\mathrm{zz}$ as
well as a generic resonance offset Hamiltonian
$\mathcal{H}_0^\mathrm{z}$, which describes inhomogeneous
resonance broadening caused for example by chemical shifts or by a
static quadrupolar interaction:
\begin{equation}
    \begin{split}
        \mathcal{H}&=\mathcal{H}_0^\mathrm{z}+\mathcal{H}_\mathrm{d}^\mathrm{zz}\\
        &=h\sum_i\Delta\nu_iI_{\mathrm{z},i}+h\sum_{i<j}\nu_{ij}\left(3I_{\mathrm{z},i}I_{\mathrm{z},j}-\mathbf{I}_i\mathbf{\cdot}\mathbf{I}_j\right)\;\mathrm{,}\label{Eq1}
    \end{split}
\end{equation}
where $\Delta\nu_i$ denotes the resonance frequency offset of the
$i$-th nuclear spin and $\nu_{ij}$ is the dipolar coupling
constant between two spins $I_i$ and $I_j$. The free decay of
transverse magnetisation under this Hamiltonian is described by a
rate $\Gamma\propto 1/T_2^*$.

\begin{figure}[htbp!]
    \includegraphics[width=0.6\linewidth]{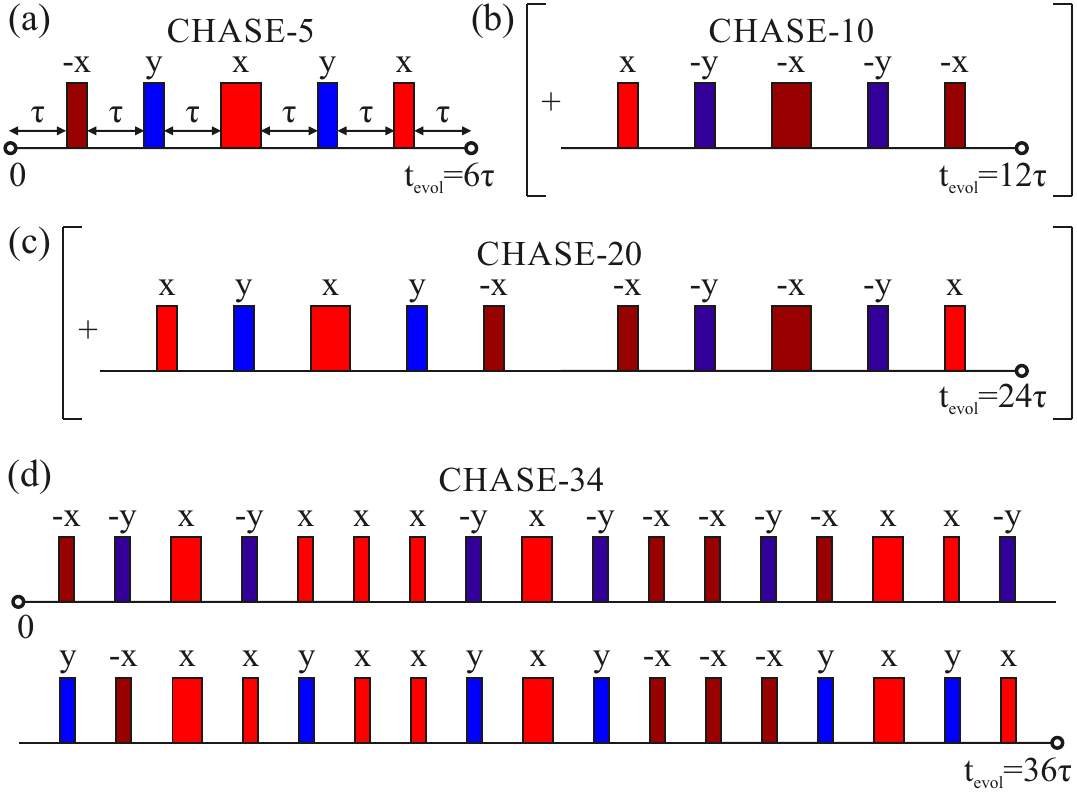}
    \caption{\label{Fig1} Pulse sequences for refocusing inhomogeneous
        and dipolar broadening. The rf carrier phases are
        $\varphi_\mathrm{x}=0$, $\varphi_{-\mathrm{x}}=\;\pi$ for $\pm$x
        rotations around the $\hat{e}_\mathrm{x}$ axis, and
        $\varphi_\mathrm{y}=\pi/2$, $\varphi_{-\mathrm{y}}=3\pi/2$ for
        $\pm$y rotations around $\hat{e}_\mathrm{y}$. Narrow pulses
        indicate $\pi/2$-rotations with pulse time $t_\pi/2$ and broad
        pulses correspondingly represent $\pi$-rotations with duration
        $t_\pi$. (a) The combined Hahn and solid echo sequence CHASE-5
        with cycle time $t_\mathrm{c}=3t_\pi+\tau_\mathrm{evol}$
        consisting of the total gate time $3t_\pi$ and total free
        evolution time $\tau_\mathrm{evol}=6\tau$ (where $\tau\gg t_\pi$
        is the free evolution time between two pulses). (b) Extension to
        CHASE-10, which is less sensitive to finite pulse durations
        $t_\pi>0$. (c) Using symmetry considerations, a further optimised
        sequence CHASE-20 is constructed. (d) The longest sequence
        CHASE-34 has the best refocusing capability for
        $t_\pi\rightarrow0$.}
\end{figure}

Using AHT as a benchmark tool (see details in Supplemental
Material \cite{supplementary}) we have analysed various
combinations of $\pi$ and $\pi/2$ pulses to find those that
maximise the spin bath coherence while minimising the pulse
sequence length. The shortest efficient cycle which gives a
noticeable increase of bath coherence (CHASE-5) contains only 5
pulses and is illustrated in Fig. \hyperref[Fig1]{1(a)}. Assuming
infinitely short pulses ($t_\pi\rightarrow0$), the zeroth order
average Hamiltonian $\propto\Gamma$ vanishes. The leading residual
contribution to dephasing is a first order term $\propto \hbar
t_\mathrm{c}\Gamma^2$ mixing contributions from the inhomogeneous
broadening Hamiltonian and the dipolar interaction
\cite{rhim1973b,burum1979}:
\begin{equation}
    \bar{\mathcal{H}}_\mathrm{CHASE-5}=\frac{\mathrm{i}t_\mathrm{c}}{18\hbar}[\mathcal{H}_\mathrm{d}^\mathrm{zz}-\mathcal{H}_\mathrm{d}^\mathrm{xx},\mathcal{H}_0^\mathrm{y}]+\mathcal{O}(\hbar t_\mathrm{c}^2\Gamma^3)\;\mathrm{,}\label{Eq2}
\end{equation}
where $t_\mathrm{c}$ is the full cycle time and
$\mathcal{H}_\mathrm{d}^\mathrm{xx}$, $\mathcal{H}_0^\mathrm{y}$
are the dipolar and inhomogeneous broadening Hamiltonians acting
along orthogonal equatorial axes $\hat{e}_\mathrm{x}$ and
$\hat{e}_\mathrm{y}$ (see Supplemental Material
\cite{supplementary} for the full definition).

Under realistic experimental conditions, the assumption of infinitely short pulses is often not justified. For finite pulse durations $t_\pi$, the zeroth order average Hamiltonian does not vanish under CHASE-5, reducing the capability of the cycle to increase the spin bath coherence. However, we can obtain an average Hamiltonian of the form of Eq. \ref{Eq2} even for finite $t_\pi$ by extending our cycle to CHASE-10 as illustrated in Fig. \hyperref[Fig1]{1(b)}, analogous to the pure solid echo extension from WAHUHA to MREV \cite{rhim1973,mansfield1973}. Furthermore, by adding the pulse block shown in Fig. \hyperref[Fig1]{1(c)} we can symmetrise the cycle to CHASE-20 and remove the first-order mixing term, condensing the average Hamiltonian to $\bar{\mathcal{H}}_\mathrm{CHASE-20}\propto\mathcal{O}(\hbar t_\mathrm{c}^2\Gamma^3)$ independent of the pulse duration $t_\pi$.

Finally, we identify a longer sequence CHASE-34 (see Fig.
\hyperref[Fig1]{1(d)}) with a total gate time $20t_\pi$ which
reduces the average Hamiltonian to a second order mixing term for
$t_\pi\rightarrow0$ but has non-vanishing lower-order terms for
finite pulse durations. A comprehensive overview of the AHT
calculations and residual Hamiltonians for the sequences shown in
figure \ref{Fig1} can be found in the Supplemental Material
\cite{supplementary}.

We study the performance of these sequences experimentally on
individual charge-free InGaAs QDs with $\sim10^5$ nuclear spins.
Here, we follow the ODNMR pump-probe scheme used in
Ref.~\cite{chekhovich2015}: the QD sample is kept at $T=4.2$~K and
is subjected to a strong magnetic field of $B_\mathrm{z}=8$~T.
Using a confocal microscopy setup in Faraday orientation, we
prepare the nuclear spin bath optically through
polarisation-selective pumping of an exciton transition (dynamic
nuclear polarisation, DNP). In this way, we can achieve
hyperfine-mediated spin bath polarisation degrees of up to $65\%$
\cite{eble2006,puebla2013}. Radio frequency (rf) fields are
coupled to the QD via a multi-winding copper coil in close
proximity to the sample. Changes in the final bath polarisation
are probed with a weak optical pulse measuring the splitting of
the neutral exciton Zeeman doublet\cite{chekhovich2012}.

We perform resonant pulsed NMR measurements on the inhomogeneously
broadened central spin transition $-1/2\leftrightarrow+1/2$ of the
$^{75}$As (inhomogeneous width of
$\Delta\nu_\mathrm{inh}\sim40$~kHz) and $^{71}$Ga
($\Delta\nu_\mathrm{inh}\sim10$~kHz) nuclear spin ensembles
\cite{chekhovich2012,chekhovich2015}. The phases of the
$\pi$-pulses of all sequences are chosen to produce spin rotations
around the $\hat{e}_\mathrm{x}$ axis of the rotating frame. In
each experiment, a $\pi/2$-pulse is applied prior to the
multipulse cycle to initialise the spin state. We conduct
experiments with initial $\pi/2$ rotation around the
$\hat{e}_\mathrm{x}$ axis (Carr-Purcell or CP-like sequences
\cite{carr1954}, denoted as `-X') and around the
$\hat{e}_\mathrm{y}$ axis (Carr-Purcell-Meiboom-Gill or CPMG-like
sequences \cite{meiboom1958}, `-Y'): in this way we distinguish
between a genuine improvement of the spin coherence and spin
locking effects \cite{suh1994,dementyev2003,watanabe2003}, which
only stabilize spin magnetization along a certain direction. A
final $\pi/2$-pulse is an inverse of the initialisation pulse and
projects the refocused magnetisation along the
$\hat{e}_\mathrm{z}$ axis for optical
readout~\cite{chekhovich2015}.

\begin{figure}[htbp!]
    \includegraphics[width=0.7\linewidth]{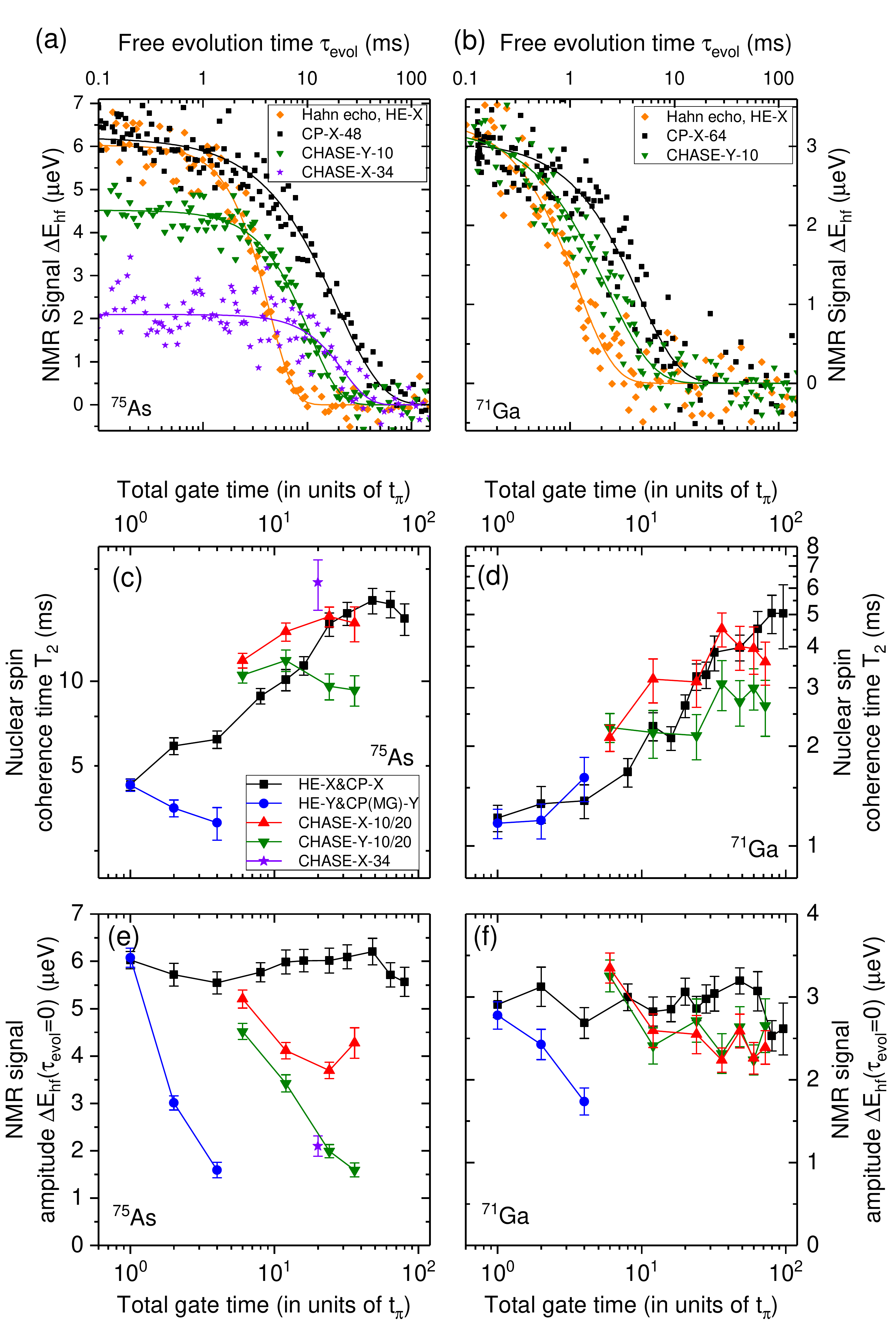}
    \caption{\label{Fig2} Nuclear spin polarisation decay under pulsed
        control of the $^{75}$As and $^{71}$Ga spin ensembles in a single
        InGaAs QD. (a,b) Decay of the polarisation $\Delta E_\mathrm{hf}$
        as a function of the total free evolution time
        $\tau_\mathrm{evol}$ for different control sequences. Symbols mark
        experimental data and solid lines show best fits with Eq.
        \ref{Eq3}. (c,d) Dependence of the fitted nuclear spin coherence
        time $T_2$ on the number of sequence cycles $n$ expressed as total
        rf gate time (in units of the $\pi$-pulse time $t_\pi$). Error
        bars mark $90\%$ confidence intervals. (e,f) Respective fitted
        echo amplitude $\Delta E_\mathrm{hf}(\tau_\mathrm{evol}=0)$ as a
        function of the total gate time. The data for one cycle of HE and
        CHASE-10 is combined with the data for integer cycle numbers $n$
        of CP and CHASE-20, respectively.}
\end{figure}

Representative data for experiments on $^{75}$As and $^{71}$Ga is
shown by the solid symbols in Figs \hyperref[Fig2]{2(a)} and
\hyperref[Fig2]{2(b)}, respectively. The values of the nuclear
spin coherence times $T_2$ and echo amplitudes $\Delta
E_\mathrm{hf}(\tau_\mathrm{evol}=0)$ in Figs~\hyperref[Fig2]{2(c-f)} are obtained by fitting the data with a
compressed exponential decay function
\begin{equation}
    \Delta E_\mathrm{hf}(\tau_\mathrm{evol})=\Delta E_\mathrm{hf}(\tau_\mathrm{evol}=0)\cdot\text{e}^{-(\tau_\mathrm{evol}/T_2)^\beta}\;\text{,} \label{Eq3}
\end{equation}
where $\Delta E_\mathrm{hf}$ denotes the change in the measured
hyperfine shift due to rf-induced depolarisation of the nuclear
spin bath, $\tau_\mathrm{evol}$ is the total free evolution time
over $n$ pulse cycles, $\beta\in[1,2]$ is a compression
factor~\cite{bargill2012}, $T_2$ describes decoherence of the spin
bath during free evolution, while reduction of the echo amplitude
$\Delta E_\mathrm{hf}(\tau_\mathrm{evol}=0)$ compared to the
initial magnetization $\Delta E_\mathrm{hf}(t=0)$ quantifies the
imperfections of pulse spin rotations.

In order to analyse the influence of spin locking effects
\cite{suh1994,dementyev2003,watanabe2003}, we test a series of
CP-X and \mbox{CP(MG)-Y} sequences with alternating pulse carrier
phase (sequence cycle
$-\tau/2-\pi_\mathrm{x}-\tau-\pi_\mathrm{-x}-\tau/2-$). With
increasing number of cycles $n$, expressed in terms of the total
rf gate time, we observe a strong increase of the measured $T_2$
under CP-X for both isotopes (black squares in
Figs~\hyperref[Fig2]{2(c,d)}). By contrast, no significant
increase of $T_2$ is observed under CP-Y (blue circles). However,
the CP-Y echo amplitude is rapidly reduced with increasing $n$
(Figs \hyperref[Fig2]{2(e,f)}), owing to the limited available rf
power resulting in violation of the `hard pulse' condition
\cite{supplementary}.

The contrasting behaviour of $T_2$ under alternating phase CP-X/Y
has been observed in other systems
\cite{dementyev2003,watanabe2003,ridge2014} and has been
attributed variably to spin locking \cite{li2008,ridge2014} or
stimulated echoes \cite{franzoni2005}. Here, we ascribe the
up-to-fourfold increase of $T_2$ under CP-X to a form of pulsed
spin locking arising from dipolar evolution during the finite
$\pi$-pulse duration \cite{li2008}: Our interpretation is based on
observation that the spin lock disappears for small pulse-to-cycle
time ratios $t_\pi/t_\mathrm{c}$ (see Supplemental Material
\cite{supplementary}).

We now examine the spin bath coherence under CHASE-10/20
sequences. In order to account for the spin locking effects
discussed above we conduct all experiments with both -X and -Y
initialization. In experiments on $^{75}$As, a marked increase of
$T_2^\mathrm{CHASE-Y-10}=10.5\pm0.7$~ms compared to the Hahn echo
decay $T_2^\mathrm{HE-X}=4.3\pm0.2$~ms is observed even under a
single cycle of CHASE-Y-10 (green triangles in Fig. \ref{Fig2}).
We find a similar proportional increase from
$T_2^\mathrm{HE-X}=1.2\pm0.1$~ms to
$T_2^\mathrm{CHASE-Y-10}=2.3\pm0.2$~ms for $^{71}$Ga. In
comparison, the additional coherence gain under CHASE-Y-20 is only
marginal. However, as shown in Figs \hyperref[Fig2]{2(e,f)}, the
preservation of the echo amplitude $\Delta
E_\mathrm{hf}(\tau_\mathrm{evol}=0)$ under CHASE-Y-20 is more
robust compared to CP-Y. The CHASE-X-10/20 coherence time (red
triangles in Figs. \hyperref[Fig2]{2(c,d)}) exceeds the $T_2$ of
CHASE-Y-10/20, suggesting the presence of spin-locking under
CHASE-X-10/20. Thus we use CHASE-Y-10/20 to examine the spectral
properties of the spin bath dynamics in a dynamical decoupling
fashion: We observe no further coherence gain under up to $n=6$
cycles of CHASE-Y-20, suggesting that environment noise (e.g.
produced by charge fluctuations) is negligible over a broad
frequency domain of up to $\sim100$~kHz (given by an average
inter-pulse delay of $\sim10$~$\mu$s), as expected for neutral QDs
\cite{cywinski2008,chekhovich2015}. Recent work has shown,
however, that the nuclear spin bath coherence is drastically
reduced when the QD is occupied by an electron \cite{wust2016}. We
expect that CHASE sequences will be suited to restore the bath
coherence, offering a pathway for electron spin manipulation in a
quiescent QD environment.

Finally, we present an experiment using CHASE-34 (violet stars in
Fig. \ref{Fig2}). We find that only the CP-like cycle yields a
measurable decay curve for $^{75}$As whereas the $^{71}$Ga echo
amplitude $\Delta E_\mathrm{hf}(\tau_\mathrm{evol}=0)$ is too
small to be resolved. Tantalisingly, we observe
$T_2^\mathrm{CHASE-X-34}=22.4\pm4.5$~ms (violet stars in Fig.
\hyperref[Fig2]{2(a)}) -- nearly a fivefold increase of the bath
coherence time compared to HE. Numerical simulations (see
Supplemental Material \cite{supplementary}) indicate that the
reduction of $\Delta E_\mathrm{hf}(\tau_\mathrm{evol}=0)$ in this
34-pulse cycle is not due to the limited rf pulse bandwidth, but
is likely related to small pulse calibration errors, offering in
principle a route for further improvements.

\begin{figure}[htbp!]
    \includegraphics[width=0.8\linewidth]{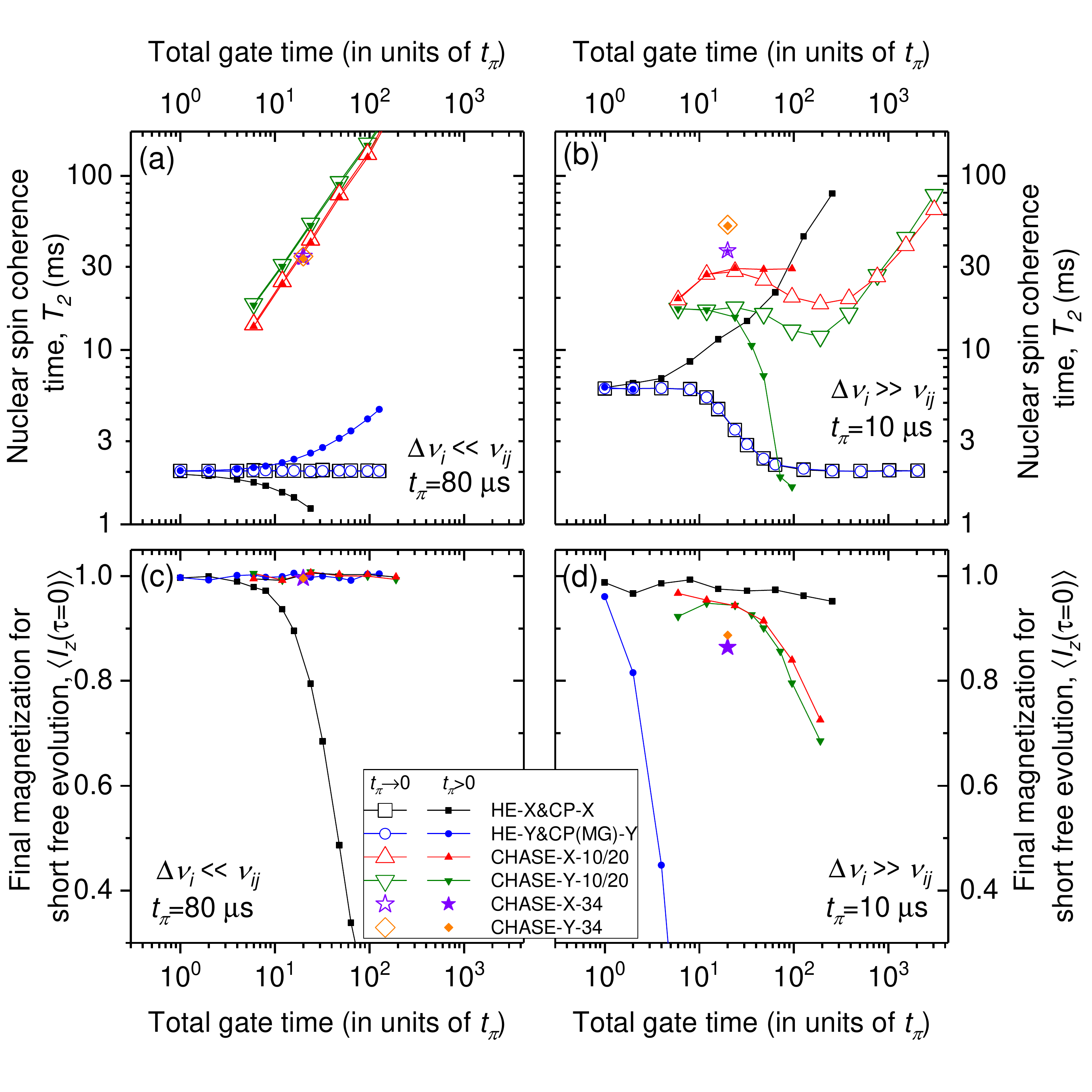}
    \caption{\label{Fig3} Simulated evolution of a dipolar
        coupled ensemble of twelve $^{75}$As spins under control sequences
        (a,c) with small and (b,d) large inhomogeneous (quadrupolar)
        broadening $\Delta\nu_i$. Figs (a) and (b) show the respective
        fitted nuclear spin coherence times $T_2$ as a function of the
        total gate time. The gate time dependence of the corresponding
        echo amplitudes for short free evolution $\braket{I_z(0)}$ is
        shown in (c) and (d). Simulations are done for both infinitely
        sharp ($t_\pi=0$, open symbols) and finite pulses (solid symbols),
        where we set $t_\pi=80\;\mu\text{s}$ for $\Delta\nu_i\ll\nu_{ij}$
        and use $t_\pi=10\;\mu\text{s}$ in the limit of
        $\Delta\nu_i\gg\nu_{ij}$. Further simulation details and
        representative decay curves are shown in the Supplemental Material
        \cite{supplementary}.}
\end{figure}

The experimental observations are corroborated by extensive first
principle quantum mechanical simulations of the nuclear spin bath
evolution under the studied pulse sequences. We consider an
ensemble of twelve dipolar coupled $^{75}$As spins and study the
evolution of the resonantly driven central transition in the
limits of large ($\Delta\nu_i\gg\nu_{ij}$) and vanishing
($\Delta\nu_i\ll\nu_{ij}$) inhomogeneous broadening. In this way,
we can confirm the experimental results in self-assembled QDs with
strong static quadrupolar broadening and explore the applicability
of CHASE sequences to more homogeneous dilute spin systems as
encountered in defect centres in diamond or donors in silicon. In
addition, we study the influence of finite pulse durations on the
bath evolution by testing our sequences with both infinitely short
pulses $t_\pi\rightarrow0$ and realistic pulse durations
$t_\pi\sim10-100\;\mu\text{s}$.

Figure \hyperref[Fig3]{3} shows the fitted coherence times (a,b)
and echo amplitudes (c,d) for simulated decay curves in case of
negligible inhomogeneous resonance broadening
$\Delta\nu_i\ll\nu_{ij}$ (a,c) and large broadening
$\Delta\nu_i\gg\nu_{ij}$ (b,d). Simulation data at infinitely
short (finite) pulses is marked by open (solid) symbols. The
simulations with $\Delta\nu_i\gg\nu_{ij}$ and finite pulses are in
very good agreement with the experiments (cf.
Fig.~\hyperref[Fig2]{2(c-f)}): namely, the increase in $T_2$ with
increasing number of cycles $n$ under CP-X (spin locking), the
reduction of echo amplitude with growing $n$ under CP-Y, and the
extension of $T_2$ under CHASE-10/20 are all well reproduced.
Moreover, the CHASE-10/20 echo amplitudes are stable under
increasing $n$, demonstrating the superior performance of CHASE
under `soft pulse' conditions compared to previously introduced
sequences \cite{maurer2012,moiseev2015,moiseev2015b} (see
additional simulations in Supplemental Material
\cite{supplementary}).

Having established the validity of the simulations we apply them
to the regimes not accessible in experiments on self-assembled
QDs. In the case of $\Delta\nu_i\ll\nu_{ij}$ applicable e.g. to
defect spins in diamond and SiC, the effect of the CHASE sequences
is analogous to that of a solid echo sequence
\cite{waugh1968,rhim1973,mansfield1973}: dipolar broadening can be
suppressed arbitrarily well with increasing $n$ resulting in
increase of $T_2$ (triangles in Fig.~\hyperref[Fig3]{3(a)}).
Importantly, the CHASE-10/20 echo amplitudes
(Fig.~\hyperref[Fig3]{3(c)}) remain close to ideal $\approx1$
under finite ('soft') pulses even at large $n$: there is a
potential in applying CHASE-10/20 to electron spin qubits for
dynamical decoupling from the nuclear spin bath and for
simultaneous suppression of the decoherence that arises
from qubit-qubit dipolar interactions
\cite{tyryshkin2012,delange2012} and can not be handled by the
standard $\pi$-pulse sequences.

Finally, we make a note on some peculiarities predicted in the
simulations. For large inhomogeneity ($\Delta\nu_i\gg\nu_{ij}$)
and short pulses ($t_\pi\rightarrow0$) the CP-X/Y coherence time
$T_2$ decreases with increasing $n$ (open squares and circles in
Fig.~\hyperref[Fig3]{3(b)}) and asymptotically approaches the
$T_2$ value obtained for weak inhomogeneity
($\Delta\nu_i\ll\nu_{ij}$) (Fig.~\hyperref[Fig3]{3(a)}). This is
unexpected, since $\pi$-pulse trains do not modify the dipolar
Hamiltonian. We tentatively ascribe this to fast spin rotations
induced by the infinitely short rf pulses: such rotations
effectively shorten the spin lifetime, broaden the homogeneous NMR
linewidths and re-enable the dipolar nuclear spin flip-flops which
are otherwise inhibited by inhomogeneous broadening. Note that
similar trends are observed in the simulations with CHASE-10/20
sequences, where the competing effects of the re-enabled
flip-flops and convergence of the average Hamiltonian lead to a
non-monotonic dependence of $T_2$ on $n$. We also point out that
under small inhomogeneous broadening $\Delta\nu_i\ll\nu_{ij}$ and
finite pulse durations, the increase (decrease) of $T_2$ under
CP-Y (CP-X) observed in Fig. \hyperref[Fig3]{3(a)} is reversed
compared to the $\Delta\nu_i\gg\nu_{ij}$ case (Fig.
\hyperref[Fig3]{3(b)}) -- an unexpected result requiring further
investigation.

In summary, we have introduced a set of multiple pulse cycles
which allow efficient simultaneous refocusing of inhomogeneous and
dipolar broadening. Beyond the current work, these sequences may
have potential to restore the spin bath coherence in charged QDs
which have recently been shown to possess far shorter Hahn echo
decay times \cite{wust2016}. In this way, one would aim to create
a deterministically evolving spin environment for a central
electron spin. In addition, CHASE could be used to enhance the
coherence of defect centres and dopants beyond the limits of
standard dynamical decoupling protocols as the parasitic
decoherence channel of instantaneous diffusion
\cite{delange2012,tyryshkin2012} is quenched. Further optimisation
of spin bath fluctuation freezing can be explored using techniques
such as optimal control \cite{khaneja2005}.

The authors are grateful to A. I. Tartakovskii for useful
discussions. This work was supported by the EPSRC Programme Grant
EP/J007544/1, ITN S\textsuperscript{3}NANO. E.A.C. was supported
by a University of Sheffield Vice-Chancellor's Fellowship and a
Royal Society University Research Fellowship. Computational
resources were provided in part by the University of Sheffield HPC
cluster Iceberg.


\renewcommand{\thesection}{Supplemental Note \arabic{section}}
\setcounter{section}{0}
\renewcommand{\thefigure}{\arabic{figure}}
\renewcommand{\figurename}{Supplemental Figure}
\setcounter{figure}{0}
\renewcommand{\theequation}{\arabic{equation}}
\setcounter{equation}{0}
\renewcommand{\thetable}{\arabic{table}}
\renewcommand{\tablename}{Supplemental Table}
\setcounter{table}{0}

\renewcommand{\citenumfont}[1]{S#1}
\makeatletter
\renewcommand{\@biblabel}[1]{S#1.}
\makeatother

\pagebreak \pagenumbering{arabic}

\section*{Supplemental Material}

The supplemental material
contains further details on the theoretical tools used to design
the CHASE pulse sequences presented in the main text, an analysis
of the influence of pulsed spin locking and a discussion of the
respective influence of pulse parameters on the performance of
these spin control sequences. Finally, the methods used to
simulate the spin bath evolution are described with additional
results for related pulse sequences from literature.


\section{\label{sec:sn1}Average Hamiltonian theory}

Average Hamiltonian theory (AHT) is an established tool for the
theoretical characterisation and analysis of pulse sequences for
magnetic resonance spin
control\cite{haeberlen1968s,rhim1973bs,burum1979s}. Within certain
constraints it allows the time evolution of a given spin
Hamiltonian under interaction with a periodic time-dependent
external magnetic field to be approximated. We use AHT to
determine how well a frequency offset Hamiltonian
$\mathcal{H}_0^\mathrm{z}$ and a dipolar coupling term
$\mathcal{H}_\mathrm{d}^\mathrm{zz}$ can be suppressed
simultaneously by the CHASE sequences introduced in the main text.

To this end, we consider a nuclear spin ensemble $I_i$ with spin
$1/2$. The evolution of the
wavefunction $\psi(t)$ describing the state of the nuclear spin
bath is determined by the Schr\"{o}dinger equation:
\begin{gather}
\partial\psi(t)/\partial t=-(\textrm{i}/\hbar)\mathcal{H}(t)\psi(t)\;\mathrm{,}\label{SEq1a}\\
\mathcal{H}(t) = \mathcal{H}_\textrm{L}^\mathrm{z} + \mathcal{H}_\textrm{0}^\mathrm{z} +\mathcal{H}_\textrm{d}^\mathrm{zz} + \mathcal{H}_\textrm{rf}(t)\;\mathrm{,}\label{SEq1b}
\end{gather}
where the Hamiltonian $\mathcal{H}(t)$ is the sum of the Larmor
term $\hat H_\textrm{L}$ describing interaction of the spins with
a static magnetic field $B_\mathrm{z}$ along the
$\hat{e}_\mathrm{z}$ axis, the offset term
$\mathcal{H}_0^\mathrm{z}$ describing static resonance frequency
shifts, the dipolar term $\mathcal{H}_\mathrm{d}^\mathrm{zz}$
describing nuclear-nuclear spin interaction and the
radio-frequency (rf) term $\mathcal{H}_\textrm{rf}(t)$ describing
the effect of the oscillating magnetic field inducing nuclear
magnetic resonance.

We use transformation into the frame rotating around the direction
of the static magnetic field ($\hat{e}_\mathrm{z}$ axis) at the
radio-frequency. In this way the effect of the static magnetic
field is eliminated ($\mathcal{H}_\textrm{L}^\mathrm{z}=0$) and the oscillating
rf field becomes static (see Section 5.5 in \cite{slichter1996s}).
The explicit time dependence in $\mathcal{H}_\textrm{rf}(t)$ is
then only due to the pulsed nature of the rf field.

The individual terms are explicitly defined as
\begin{gather}
\mathcal{H}_0^\mathrm{z}=h\sum_i\Delta\nu_iI_{\mathrm{z},i}\;\mathrm{,}\label{SEq2}\\
\mathcal{H}_\mathrm{d}^\mathrm{zz}=h\sum_{i<j}\nu_{ij}\left(3I_{\mathrm{z},i}I_{\mathrm{z},j}-\mathbf{I}_i\mathbf{\cdot}\mathbf{I}_j\right)\;\mathrm{,}\label{SEq3}\\
\mathcal{H}_\mathrm{rf}(t)=-h\nu_\mathrm{rf}(t)\sum_iI_{\varphi,i}\;\mathrm{.}\label{SEq4}
\end{gather}
In the studied quantum dots (QDs) the resonance offset term
$\mathcal{H}_0^\mathrm{z}$ is dominated by the static quadrupolar
frequency shifts ($\Delta\nu_i$ for the $i$-th nuclear spin),
although in other systems $\Delta\nu_i$ could also include
different static frequency offsets such as chemical shifts or
magnetic field gradients. In addition, we consider a truncated
dipolar coupling term $\mathcal{H}_\mathrm{d}^\mathrm{zz}$ with
coupling strength
\begin{equation}
\nu_{ij}=\frac{\mu_0}{4\pi}\frac{\hbar}{2\pi}\frac{\gamma^2}{2}\frac{1-3\cos^2{\theta}}{r^3}\;\mathrm{,}\label{SEq4b}
\end{equation}
between two spins $I_i$ and $I_j$. Here,
$\mu_0=4\pi\cdot10^{-7}\;\text{N/A}^2$ is the magnetic constant,
$\gamma$ is the nuclear gyromagnetic ratio, and $r$ denotes the
length of the vector connecting the two spins, which forms angle
$\theta$ with the $\hat{e}_\mathrm{z}$ axis.

Interaction with resonant rf pulses is described by the
time-dependent term $\mathcal{H}_\mathrm{rf}(t)$ where the field
amplitude $\nu_\mathrm{rf}(t)=\nu_0$ during a pulse and
$\nu_\mathrm{rf}(t)=0$ otherwise. The spin operator $I_\varphi$
determines the in-plane rotation axis about which the spin bath
precesses under the rf field with a given phase~$\varphi$:
\begin{equation}
I_\varphi=I_\mathrm{x}\cos\varphi+I_\mathrm{y}\sin\varphi\;\mathrm{.}\label{SEq5}
\end{equation}

Here, we want to study the spin bath evolution under the
'internal' static terms
$\mathcal{H}_\mathrm{int}=\mathcal{H}_0^\mathrm{z}+\mathcal{H}_\mathrm{d}^\mathrm{zz}$
in the interaction frame of $\mathcal{H}_\mathrm{rf}(t)$. AHT can
give an approximate description of the time evolution after one or
more full rf cycles if three conditions are met: (i) The rf
Hamiltonian is periodic over the cycle time $t_\mathrm{c}$, i.e.
$\mathcal{H}_\mathrm{rf}(t+t_\mathrm{c})=\mathcal{H}_\mathrm{rf}(t)$.
(ii) The net spin rotation after a full rf cycle is a multiple of
$2\pi$. (iii) Since AHT is a perturbation method in terms of
$t_\mathrm{c}/T_2^*$, the solution only converges quickly if
$|\mathcal{H}_0^\mathrm{z}|t_\mathrm{c}/\hbar\ll1$ and
$|\mathcal{H}_\mathrm{d}^\mathrm{zz}|t_\mathrm{c}/\hbar\ll1$.

We write the total time-evolution operator as
\begin{equation}
\mathcal{U}(t)=\mathcal{T}\exp\left[-\frac{\mathrm{i}}{\hbar}\int_0^tdt'\mathcal{H}(t')\right]=\mathcal{U}_\mathrm{rf}(t)\mathcal{U}_\mathrm{int}(t)\;\mathrm{,}\label{SEq6}
\end{equation}
with Dyson time-ordering operator $\mathcal{T}$ and
\begin{align}
\mathcal{U}_\mathrm{rf}(t)=\mathcal{T}\exp\left[-\frac{\mathrm{i}}{\hbar}\int_0^tdt'\mathcal{H}_\mathrm{rf}(t')\right]\;\mathrm{,}\label{SEq7} \\
\mathcal{U}_\mathrm{int}(t)=\mathcal{T}\exp\left[-\frac{\mathrm{i}}{\hbar}\int_0^tdt'\tilde{\mathcal{H}}_\mathrm{int}(t')\right]\;\mathrm{,}\label{SEq8}
\end{align}
where we introduced the toggling frame Hamiltonian
\begin{equation}
\tilde{\mathcal{H}}_\mathrm{int}(t)=\mathcal{U}_\mathrm{rf}^{-1}(t)\mathcal{H}_\mathrm{int}\mathcal{U}_\mathrm{rf}(t)\;\mathrm{.}\label{SEq9}
\end{equation}
While an exact solution for Eq. \eqref{SEq8} is generally challenging to find, an approximate description of the spin bath evolution at times $t=n\cdot t_\mathrm{c}$ can be found if conditions (i)-(iii) are fulfilled. In this case, we can apply a Magnus expansion to replace the expression of Eq. \eqref{SEq8} by an effective average Hamiltonian $\bar{\mathcal{H}}$ such that
\begin{equation}
\begin{split}
\mathcal{U}_\mathrm{int}(nt_\mathrm{c})&=\exp\left[-\frac{\mathrm{i}}{\hbar}nt_\mathrm{c}\bar{\mathcal{H}}\right]\\
&=\exp\left[-\frac{\mathrm{i}}{\hbar}nt_\mathrm{c}\left(\bar{\mathcal{H}}^{(0)}+\bar{\mathcal{H}}^{(1)}+\dotso\right)\right]\;\mathrm{,}\label{SEq10}
\end{split}
\end{equation}
with leading order terms
\begin{align}
\bar{\mathcal{H}}^{(0)}&=\frac{1}{t_\mathrm{c}}\int_0^{t_\mathrm{c}}\tilde{\mathcal{H}}(t)\mathrm{d} t\;\mathrm{,}\label{SEq11}\\
\bar{\mathcal{H}}^{(1)}&=\frac{-\mathrm{i}}{2t_\mathrm{c}\hbar}\int_0^{t_\mathrm{c}}\mathrm{d} t_2\int_0^{t_2}\mathrm{d} t_1\left[\tilde{\mathcal{H}}(t_2),\tilde{\mathcal{H}}(t_1)\right]\;\mathrm{,}\label{SEq12}\\
\begin{split}
\bar{\mathcal{H}}^{(2)}&=\frac{1}{6t_\mathrm{c}\hbar^2}\int_0^{t_\mathrm{c}}\mathrm{d} t_3\int_0^{t_3}\mathrm{d} t_2\int_0^{t_2}\mathrm{d} t_1\\
&\qquad\Big(\left[\tilde{\mathcal{H}}(t_1),\left[\tilde{\mathcal{H}}(t_2),\tilde{\mathcal{H}}(t_3)\right]\right]\\
&\qquad\qquad+\left[\tilde{\mathcal{H}}(t_3),\left[\tilde{\mathcal{H}}(t_2),\tilde{\mathcal{H}}(t_1)\right]\right]\Big)\;\mathrm{.}\label{SEq13}
\end{split}
\end{align}
The contributions of higher order AHT terms to $\bar{\mathcal{H}}$ scale as $t_\mathrm{c}^k\Gamma^{k+1}$ for $\bar{\mathcal{H}}^{(k)}$, with free decay rate $\Gamma\propto1/T_2^*$. We thus see that in the limit of $n\rightarrow\infty$ cycles and cycle time $t_\mathrm{c}\rightarrow0$, only the zeroth order term $\bar{\mathcal{H}}^{(0)}$ remains. However, in practice, higher order contributions are rarely negligible, making e.g.\ longer solid echo sequences such as MREV\cite{rhim1973s,mansfield1973s} and BR-24\cite{burum1979s} more efficient than the shorter WAHUHA cycle\cite{waugh1968s} in many applications.

We calculate the average Hamiltonian for a spin bath interacting
with a given pulse sequence as described by equations
\eqref{SEq1b}-\eqref{SEq4} using Mathematica with the
freely-available non-commutative algebra package
NCAlgebra\cite{helton2017s}. For the zeroth order average
Hamiltonian, we consider finite pulse durations where $t_\pi$ is
the time required for a $\pi$-rotation. In this case, the cycle
time $t_\mathrm{c}$ is given by the sum of pulse times and pulse
spacings $\tau$. First and second order terms are only calculated
in the limit of infinitely sharp rf pulses $t_\pi\rightarrow0$.

\begin{table}[htb]
    \caption{Overview of AHT terms up to second order calculated for the CHASE sequences presented in the main text. Zeroth order terms are calculated assuming finite pulse duration $t_\pi$ whereas $t_\pi\rightarrow0$ is assumed for higher order terms. Unlisted terms $\bar{\mathcal{H}}_{0}^{(0),(1),(2)}$, $\bar{\mathcal{H}}_\mathrm{d}^{(0),(1),(2)}$,    $\bar{\mathcal{H}}_\mathrm{d0}^{(1),(2)}$ are zero. \label{tab1}}
    \begin{ruledtabular}
        \begin{tabular}{c|c}
            AHT term & CHASE-5 \\
            \hline
            $\bar{\mathcal{H}}_{0}^{(0)}$ & $\frac{2t_\pi}{\pi t_\mathrm{c}}\mathcal{H}_0^\mathrm{z}$ \\
            $\bar{\mathcal{H}}_\mathrm{d}^{(0)}$ & $\frac{\mathrm{i}t_\pi}{\pi t_c}\{[\mathcal{H}_\mathrm{d}^\mathrm{xx},I_\mathrm{z}]+[\mathcal{H}_\mathrm{d}^\mathrm{yy},I_\mathrm{x}-I_\mathrm{z}]-[\mathcal{H}_\mathrm{d}^\mathrm{zz},I_\mathrm{x}]\}$ \\
            \hline
            $\bar{\mathcal{H}}_\mathrm{d0}^{(1)}$ & $\frac{\mathrm{i}t_\mathrm{c}}{18\hbar}[\mathcal{H}_\mathrm{d}^\mathrm{zz}-\mathcal{H}_\mathrm{d}^\mathrm{xx},\mathcal{H}_0^\mathrm{y}]$ \\
            \hline
            $\bar{\mathcal{H}}_\mathrm{d}^{(2)}$ & $\frac{1}{2}\left(\frac{t_\mathrm{c}}{18\hbar}\right)^2[\mathcal{H}_\mathrm{d}^\mathrm{zz}-\mathcal{H}_\mathrm{d}^\mathrm{xx},[\mathcal{H}_\mathrm{d}^\mathrm{zz},\mathcal{H}_\mathrm{d}^\mathrm{yy}]]$ \\
            $\bar{\mathcal{H}}_\mathrm{d0}^{(2)}$ & $3\left(\frac{t_\mathrm{c}}{18\hbar}\right)^2\{[\mathcal{H}_0^\mathrm{y},[\mathcal{H}_0^\mathrm{y},\mathcal{H}_\mathrm{d}^\mathrm{xx}-\mathcal{H}_\mathrm{d}^\mathrm{zz}]]+[\mathcal{H}_\mathrm{d}^\mathrm{zz}-\mathcal{H}_\mathrm{d}^\mathrm{yy},[\mathcal{H}_0^\mathrm{z},\mathcal{H}_0^\mathrm{y}]]\}$ \\
            \hline
            \hline
            & CHASE-10 \\
            \hline
            $\bar{\mathcal{H}}_\mathrm{d0}^{(1)}$ & $\frac{\mathrm{i}t_\mathrm{c}}{36\hbar}[\mathcal{H}_\mathrm{d}^\mathrm{zz}-\mathcal{H}_\mathrm{d}^\mathrm{xx},\mathcal{H}_0^\mathrm{y}]$ \\
            \hline
            $\bar{\mathcal{H}}_\mathrm{d}^{(2)}$ & $\frac{1}{2}\left(\frac{t_\mathrm{c}}{36\hbar}\right)^2[\mathcal{H}_\mathrm{d}^\mathrm{zz}-\mathcal{H}_\mathrm{d}^\mathrm{xx},[\mathcal{H}_\mathrm{d}^\mathrm{zz},\mathcal{H}_\mathrm{d}^\mathrm{yy}]]$ \\
            $\bar{\mathcal{H}}_\mathrm{d0}^{(2)}$ & $3\left(\frac{t_\mathrm{c}}{36\hbar}\right)^2[\mathcal{H}_0^\mathrm{y},[\mathcal{H}_0^\mathrm{y},\mathcal{H}_\mathrm{d}^\mathrm{xx}-\mathcal{H}_\mathrm{d}^\mathrm{zz}]]$ \\
            \hline
            \hline
            &  CHASE-20 \\
            \hline
            $\bar{\mathcal{H}}_\mathrm{d}^{(2)}$ & $\frac{1}{2}\left(\frac{t_\mathrm{c}}{72\hbar}\right)^2[\mathcal{H}_\mathrm{d}^\mathrm{zz}-\mathcal{H}_\mathrm{d}^\mathrm{xx},[\mathcal{H}_\mathrm{d}^\mathrm{zz},\mathcal{H}_\mathrm{d}^\mathrm{yy}]]$ \\
            $\bar{\mathcal{H}}_\mathrm{d0}^{(2)}$ & $3\left(\frac{t_\mathrm{c}}{72\hbar}\right)^2[\mathcal{H}_0^\mathrm{y},[\mathcal{H}_0^\mathrm{y},\mathcal{H}_\mathrm{d}^\mathrm{xx}-\mathcal{H}_\mathrm{d}^\mathrm{zz}]]$ \\
            \hline
            \hline
            & CHASE-34 \\
            \hline
            $\bar{\mathcal{H}}_\mathrm{d}^{(0)}$ & $\frac{\mathrm{i}t_\pi}{\pi t_c}\{4[\mathcal{H}_\mathrm{d}^\mathrm{xx},I_\mathrm{y}-I_\mathrm{z}]+2[\mathcal{H}_\mathrm{d}^\mathrm{yy},I_\mathrm{x}+2I_\mathrm{z}]-2[\mathcal{H}_\mathrm{d}^\mathrm{zz},I_\mathrm{x}+2I_\mathrm{y}]+i\pi\mathcal{H}_\mathrm{d}^\mathrm{zz}\}$ \\
            \hline
            $\bar{\mathcal{H}}_\mathrm{d0}^{(2)}$ & $3\left(\frac{t_\mathrm{c}}{108\hbar}\right)^2\{\frac{2}{3}[\mathcal{H}_0^\mathrm{y},[\mathcal{H}_0^\mathrm{y},\mathcal{H}_\mathrm{d}^\mathrm{xx}-\mathcal{H}_\mathrm{d}^\mathrm{zz}]]+\frac{2}{3}[\mathcal{H}_\mathrm{d}^\mathrm{zz}-\mathcal{H}_\mathrm{d}^\mathrm{yy},[\mathcal{H}_0^\mathrm{z},\mathcal{H}_0^\mathrm{y}]]-\frac{1}{3}[\mathcal{H}_\mathrm{d}^\mathrm{zz}-\mathcal{H}_\mathrm{d}^\mathrm{xx},[\mathcal{H}_0^\mathrm{z},\mathcal{H}_0^\mathrm{x}]]\}$ \\
        \end{tabular}
    \end{ruledtabular}
\end{table}

The AHT terms we obtain for the CHASE sequences presented in the
main text are listed in Supplemental Table~\ref{tab1}. For
clarity, we split the $k$-th order average Hamiltonian into
contributions from the resonance offset
($\bar{\mathcal{H}}_{0}^{(k)}$) and dipolar Hamiltonian
($\bar{\mathcal{H}}_\mathrm{d}^{(k)}$). For higher orders
$k\geq1$, we also include mixed terms
($\bar{\mathcal{H}}_\mathrm{d0}^{(k)}$). The full $k$-th order
average Hamiltonian is thus given by
\begin{equation}
\bar{\mathcal{H}}^{(k)}=\bar{\mathcal{H}}_{0}^{(k)}+\bar{\mathcal{H}}_\mathrm{d}^{(k)}+\bar{\mathcal{H}}_\mathrm{d0}^{(k)}\;\mathrm{.}\label{SEq13b}
\end{equation}
Only non-vanishing terms are listed, i.e. terms which do not appear in Supplemental Table~\ref{tab1} do not contribute to the total average Hamiltonian.

The definition of the additional resonance offset and dipolar coupling Hamiltonians used in Supplemental Table~\ref{tab1} is based on equations \eqref{SEq2} and \eqref{SEq3}, i.e.
\begin{gather}
\mathcal{H}_0^\mathrm{x}=h\sum_i\Delta\nu_iI_{\mathrm{x},i}\;\mathrm{,}\qquad\mathcal{H}_0^\mathrm{y}=h\sum_i\Delta\nu_iI_{\mathrm{y},i}\;\mathrm{,}\label{SEq14}\\
\mathcal{H}_\mathrm{d}^\mathrm{xx}=h\sum_{i<j}\nu_{ij}\left(3I_{\mathrm{x},i}I_{\mathrm{x},j}-\mathbf{I}_i\mathbf{\cdot}\mathbf{I}_j\right)\;\mathrm{,}\qquad\mathcal{H}_\mathrm{d}^\mathrm{yy}=h\sum_{i<j}\nu_{ij}\left(3I_{\mathrm{y},i}I_{\mathrm{y},j}-\mathbf{I}_i\mathbf{\cdot}\mathbf{I}_j\right)\;\mathrm{.}\label{SEq15}
\end{gather}
As discussed in the main text, we see that CHASE-5 has a non-vanishing zeroth order contribution if the pulse duration $t_\pi$ is non-negligible. In order to suppress spin bath dynamics using cycles of CHASE-5, it is therefore crucial to minimise the ratio $t_\pi/t_\mathrm{c}$. The subsequent longer sequences are insensitive to finite pulse durations in zeroth order and leave progressively fewer higher order AHT terms in the $t_\pi\rightarrow0$ limit. CHASE-34 forms an exception to this behaviour. While most efficient in suppressing the spin dynamics under ideal conditions, this cycle is also prone to finite pulse effects as the zeroth order dipolar average Hamiltonian contributes to dephasing under realistic experimental conditions.

\begin{figure}[htb!]
    \includegraphics[width=0.6\linewidth]{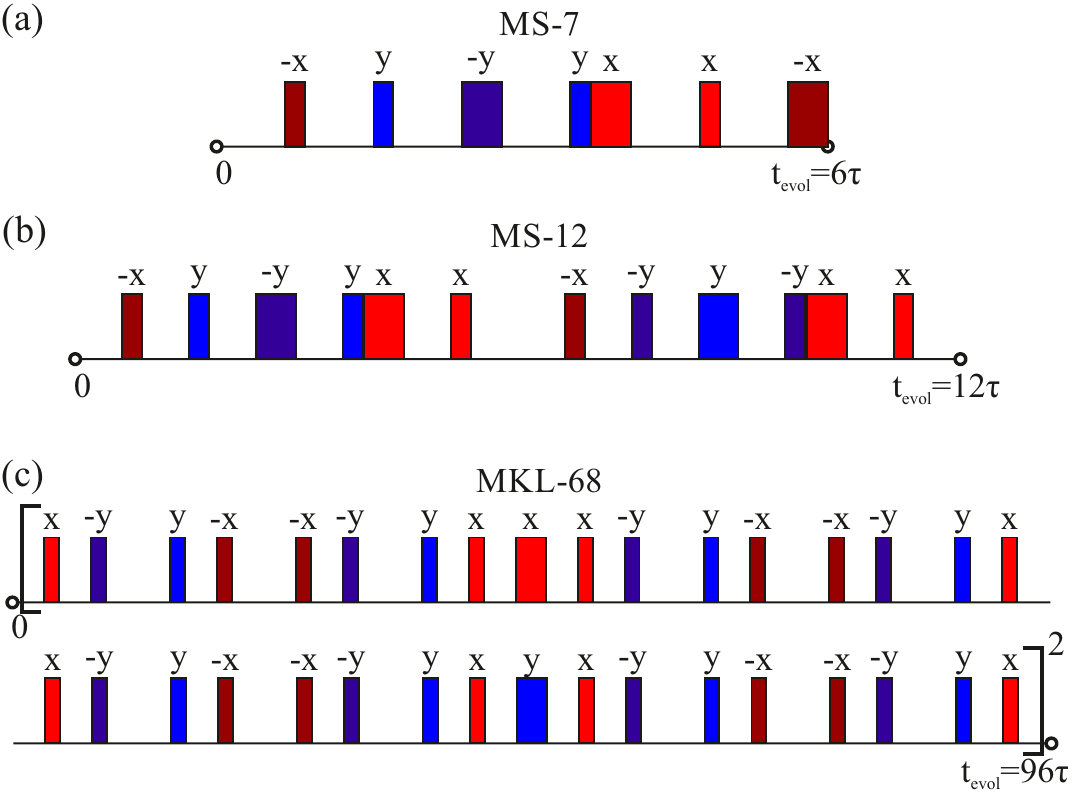}
    \caption{\label{SFig1} Schematic of alternative pulse sequences from literature for refocusing inhomogeneous and dipolar broadening. Labels are defined as in Fig. 1 of the main text. (a) MS-7 cycle as introduced by Moiseev and Skrebnev in Ref\cite{moiseev2015s}. (b) MS-12 cycle as introduced by Moiseev and Skrebnev in Ref \cite{moiseev2015bs}. (c) MKL-68 cycle as described and used by Maurer, Kucsko et al.\ in Ref \cite{maurer2012s}.}
\end{figure}

For comparison, we also calculate the AHT terms of alternative sequences from literature which have been proposed or used with the aim of suppressing both dipolar coupling and frequency offset terms. These results are listed separately in Supplemental Table~\ref{tab2}.

\begin{table}[htb]
    \caption{Overview of AHT terms calculated for other cycles referred to in the main text. \label{tab2}}
    \begin{ruledtabular}
        \begin{tabular}{c|cc}
            AHT term & MS-7 \cite{moiseev2015s} & \qquad\\
            \hline
            $\bar{\mathcal{H}}_{0}^{(0)}$ & $\frac{2t_\pi}{\pi t_\mathrm{c}}\{\mathcal{H}_0^\mathrm{y}+\mathcal{H}_0^\mathrm{z}\}$  & \qquad \\
            $\bar{\mathcal{H}}_\mathrm{d}^{(0)}$ & $\frac{\mathrm{i}t_\pi}{\pi t_c}\{[\mathcal{H}_\mathrm{d}^\mathrm{yy}-\mathcal{H}_\mathrm{d}^\mathrm{zz},I_\mathrm{x}]+\frac{5}{2}\mathrm{i}\pi\mathcal{H}_\mathrm{d}^\mathrm{xx}+\mathrm{i}\pi\mathcal{H}_\mathrm{d}^\mathrm{yy}+2\mathrm{i}\pi\mathcal{H}_\mathrm{d}^\mathrm{zz}\}$  & \qquad \\
            \hline
            $\bar{\mathcal{H}}_\mathrm{d0}^{(1)}$ & $\frac{\mathrm{i}t_\mathrm{c}}{18\hbar}[\mathcal{H}_\mathrm{d}^\mathrm{zz}-\mathcal{H}_\mathrm{d}^\mathrm{xx},\mathcal{H}_0^\mathrm{y}]$  & \qquad \\
            \hline
            $\bar{\mathcal{H}}_\mathrm{d}^{(2)}$ & $\frac{1}{2}\left(\frac{t_\mathrm{c}}{18\hbar}\right)^2[\mathcal{H}_\mathrm{d}^\mathrm{zz}-\mathcal{H}_\mathrm{d}^\mathrm{xx},[\mathcal{H}_\mathrm{d}^\mathrm{zz},\mathcal{H}_\mathrm{d}^\mathrm{yy}]]$  & \qquad \\
            $\bar{\mathcal{H}}_\mathrm{d0}^{(2)}$ & $3\left(\frac{t_\mathrm{c}}{18\hbar}\right)^2\{[\mathcal{H}_0^\mathrm{y},[\mathcal{H}_0^\mathrm{y},\mathcal{H}_\mathrm{d}^\mathrm{xx}-\mathcal{H}_\mathrm{d}^\mathrm{zz}]]+[\mathcal{H}_\mathrm{d}^\mathrm{zz}-\mathcal{H}_\mathrm{d}^\mathrm{yy},[\mathcal{H}_0^\mathrm{z},\mathcal{H}_0^\mathrm{y}]]\}$  & \qquad \\
            \hline
            \hline
            & MS-12 \cite{moiseev2015bs}  & \qquad \\
            \hline
            $\bar{\mathcal{H}}_\mathrm{d}^{(0)}$ & $\frac{\mathrm{i}t_\pi}{\pi t_c}\{[\mathcal{H}_\mathrm{d}^\mathrm{yy}-\mathcal{H}_\mathrm{d}^\mathrm{zz},2I_\mathrm{x}]+\frac{5}{2}\mathrm{i}\pi\mathcal{H}_\mathrm{d}^\mathrm{xx}+\frac{1}{2}\mathrm{i}\pi\mathcal{H}_\mathrm{d}^\mathrm{yy}+\frac{5}{2}\mathrm{i}\pi\mathcal{H}_\mathrm{d}^\mathrm{zz}\}$  & \qquad \\
            \hline
            $\bar{\mathcal{H}}_\mathrm{d}^{(2)}$ & $\frac{1}{2}\left(\frac{t_\mathrm{c}}{36\hbar}\right)^2[\mathcal{H}_\mathrm{d}^\mathrm{zz}-\mathcal{H}_\mathrm{d}^\mathrm{xx},[\mathcal{H}_\mathrm{d}^\mathrm{zz},\mathcal{H}_\mathrm{d}^\mathrm{yy}]]$  & \qquad \\
            $\bar{\mathcal{H}}_\mathrm{d0}^{(2)}$ & $3\left(\frac{t_\mathrm{c}}{36\hbar}\right)^2\{[\mathcal{H}_0^\mathrm{y},[\mathcal{H}_0^\mathrm{y},\mathcal{H}_\mathrm{d}^\mathrm{xx}-\mathcal{H}_\mathrm{d}^\mathrm{zz}]]+[\mathcal{H}_\mathrm{d}^\mathrm{zz}-\mathcal{H}_\mathrm{d}^\mathrm{yy},[\mathcal{H}_0^\mathrm{z},\mathcal{H}_0^\mathrm{y}]]\}$  & \qquad \\
            \hline
            \hline
            & MKL-68 \cite{maurer2012s} & \qquad \\
            \hline
            $\bar{\mathcal{H}}_\mathrm{d}^{(0)}$ & $\frac{t_\pi}{t_c}\{-2\mathcal{H}_\mathrm{d}^\mathrm{xx}+6\mathcal{H}_\mathrm{d}^\mathrm{yy}-\mathcal{H}_\mathrm{d}^\mathrm{zz}\}$  & \qquad \\
            \hline
            $\bar{\mathcal{H}}_\mathrm{d}^{(2)}$ & $\frac{1}{2}\left(\frac{t_\mathrm{c}}{288\hbar}\right)^2[\mathcal{H}_\mathrm{d}^\mathrm{zz}-\mathcal{H}_\mathrm{d}^\mathrm{xx},[\mathcal{H}_\mathrm{d}^\mathrm{zz},\mathcal{H}_\mathrm{d}^\mathrm{yy}]]$  & \qquad \\
            $\bar{\mathcal{H}}_\mathrm{d0}^{(2)}$ & $-3\left(\frac{t_\mathrm{c}}{288\hbar}\right)^2[\mathcal{H}_0^\mathrm{y},[\mathcal{H}_0^\mathrm{y},\mathcal{H}_\mathrm{d}^\mathrm{xx}-\mathcal{H}_\mathrm{d}^\mathrm{zz}]]$  & \qquad \\
        \end{tabular}
    \end{ruledtabular}
\end{table}

The MS-7 pulse cycle\cite{moiseev2015s} (Supplemental Fig.~\hyperref[SFig1]{1(b)}) yields average Hamiltonians identical to those of CHASE-5 in the short pulse limit. Its extension to MS-12\cite{moiseev2015bs} (Supplemental Fig.~\hyperref[SFig1]{1(c)}) removes odd-order AHT terms owing to its symmetry properties. However, unlike the longer CHASE sequences it is not robust against decoherence during a finite pulse duration $t_\pi$.

The intuitive approach of alternating MREV cycles with $\pi$-pulses (MKL-68, Supplemental Fig.~\hyperref[SFig1]{1(d)}) was employed by Maurer, Kucsko et al.\ to extend nuclear spin coherence times in diamond \cite{maurer2012s}. Again, the performance of the cycle is limited under experimental conditions by a non-vanishing zeroth-order term $\bar{\mathcal{H}}_\mathrm{d}^{(0)}$.

\clearpage

\section{\label{sec:sn2}Pulsed spin locking}
\begin{figure}[htb!]
    \includegraphics[width=0.7\linewidth]{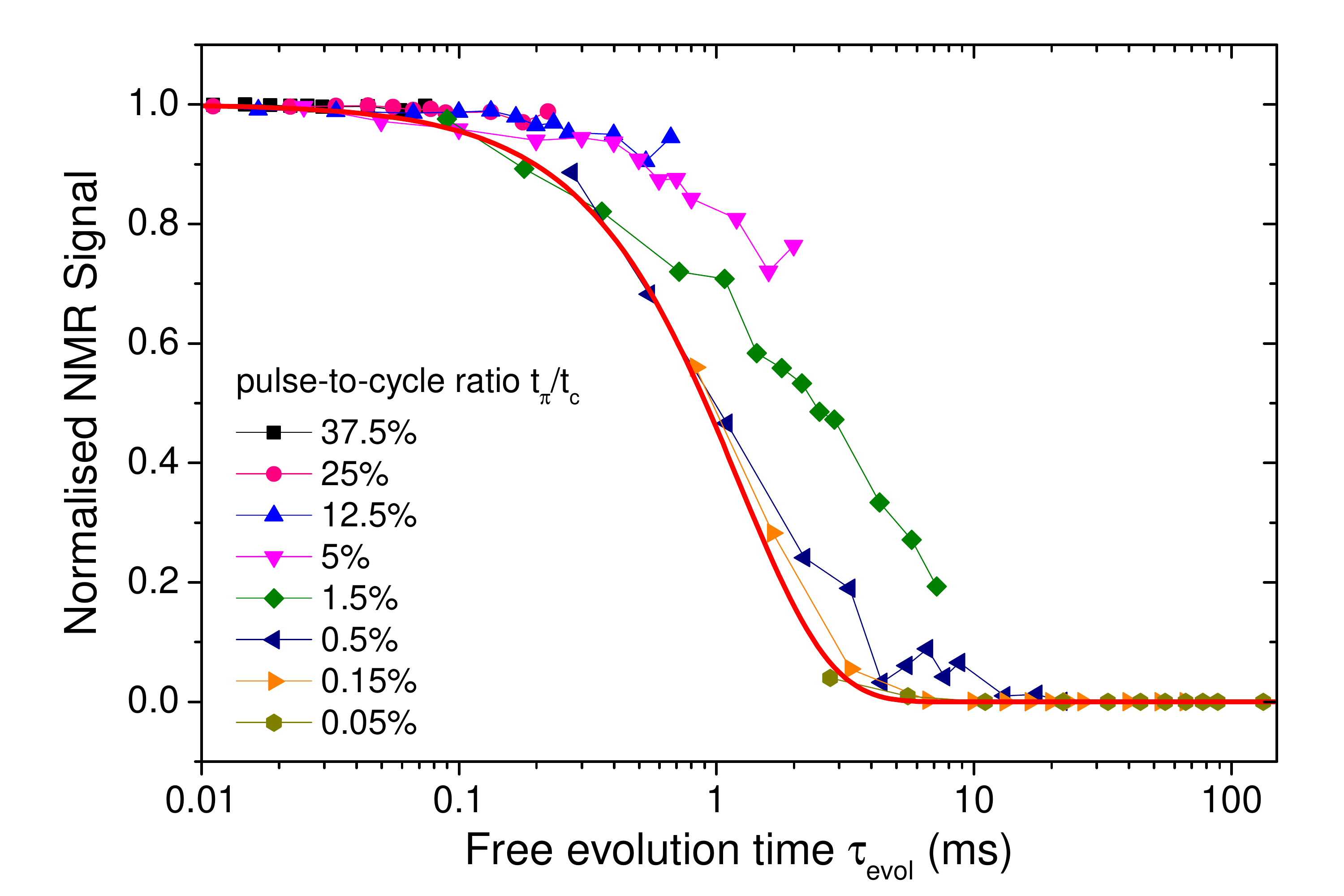}
    \caption{\label{SFig2} Dependence of the experimentally measured normalised $^{71}$Ga nuclear spin echo amplitude on the free evolution time $\tau_\mathrm{evol}$ under the phase-alternating CP-X series tested in the main text. Each trace shows data for a fixed pulse-to-cycle ratio $t_\pi/t_\mathrm{c}$ and varying $\pi$-pulse number. Values are extracted from exponential decay fits to experimental data as shown in Fig. 2(b) of the main text. The Hahn echo decay fit is shown as a solid red line for comparison.}
\end{figure}

The extended nuclear spin polarisation decay times we observe in
Figs 2(c,d) of the main text under phase-alternated
Carr-Purcell sequences (CP-X) are attributed to a form of pulsed
spin locking described theoretically by Li et al.\cite{li2008s}
This spin locking mechanism arises due to dipolar evolution during
the non-negligible $\pi$-pulse duration $t_\pi$. As shown by the
authors, in this limit application of average Hamiltonian
theory yields
\begin{equation}
\bar{\mathcal{H}}^{(0)}=\frac{4t_\pi}{\pi t_\mathrm{c}}\mathcal{H}_0^\mathrm{y}+\frac{1}{ t_\mathrm{c}}\left(4\tau\mathcal{H}_\mathrm{d}^\mathrm{zz}-t_\pi\mathcal{H}_\mathrm{d}^\mathrm{xx}\right)\;\mathrm{,}\label{eq:SEq16}
\end{equation}
for a cycle ($\tau-\pi_\mathrm{x}-2\tau-\pi_\mathrm{-x}-\tau$). The static term $\propto\mathcal{H}_0^\mathrm{y}$ in equation \eqref{eq:SEq16} is subsequently removed by transformation into a second toggling frame where we time-average over a Rabi cycle in its effective field. The twice averaged Hamiltonian is
\begin{equation}
\bar{\bar{\mathcal{H}}}^{(0)}=-\frac{1}{t_\mathrm{c}}\left(2\tau-\frac{t_\pi}{2}\right)\mathcal{H}_\mathrm{d}^\mathrm{yy}\;\mathrm{.}\label{eq:SEq17}
\end{equation}
As the initial $(\pi/2)_\mathrm{x}$ pulse of the CP-X sequence prepares the spin bath in the state $I_\mathrm{y}$ which commutes with $\mathcal{H}_\mathrm{d}^\mathrm{yy}$, the magnetisation is preserved or `locked' and no spin echo decay is predicted in zeroth order. Li et al.\ also examined the cases of fixed-phase CP-X as well as CP-Y with and without phase-alternation and found that, in agreement with our experimental results, no such effect is predicted for the CP-Y sequence tested in the current work \cite{li2008s}.

Alternative mechanisms leading to prolonged coherence times under CP-X have been suggested by other authors \cite{franzoni2005s,ridge2014s}. However, we can confidently link our experimental observation to the pulsed spin locking mechanism outlined above. A key assumption in the transition to the second toggling frame is that the spin bath evolves slowly under the Hamiltonian $\left(4\tau\mathcal{H}_\mathrm{d}^\mathrm{zz}-t_\pi\mathcal{H}_\mathrm{d}^\mathrm{xx}\right)/t_\mathrm{c}$ over the relevant timescale set by the Rabi frequency $\Omega=\tfrac{4t_\pi\Delta\nu_i}{\pi t_\mathrm{c}}$ (c.f. condition (iii) for applicability of AHT in Supplemental Note~\hyperref[sec:sn1]{1}). Hence we expect a strong dependence of the spin locking efficiency on the pulse-to-cycle time ratio $t_\pi/t_\mathrm{c}\in[0,0.5]$, where $t_\pi/t_\mathrm{c}\rightarrow0$ in the limit of infinitely short pulses and $t_\pi/t_\mathrm{c}\rightarrow0.5$ in the limit of continuous rf excitation.

In Figs 2(a) and 2(b) of the main text we show the nuclear spin polarisation decay as a function of the free evolution time for a fixed number of pulses and varying pulse spacings. In order to verify the dependence of spin locking on the pulse-to-cycle ratio, we now replot this experimental data as a function of the $\pi$-pulse number for different fixed pulse-to-cycle ratios. Supplemental Fig.~\ref{SFig2} shows that the spin locking efficiency is noticeably reduced for $t_\pi/t_\mathrm{c}\lesssim10\%$ and the Hahn echo decay (red solid line) is fully restored for $t_\pi/t_\mathrm{c}\lesssim0.5\%$.

\clearpage

\section{\label{sec:sn3} Performance beyond the `hard pulse' limit}
The central transitions (CTs) of the half-integer quadrupolar
nuclear spins in a self-assembled QD are typically inhomogeneously
broadened to $\Delta\nu_\mathrm{inh}\sim10-40$~kHz by the strain
induced electric field gradients\cite{chekhovich2012s}. In order to
perform pulse sequences on the full CT experimentally, we need to
apply rf pulses which are sufficiently broadband (`hard'), i.e.\
have a large enough amplitude to perform the desired $\pi/2$- or
$\pi$-rotation even for spins $I$ in the tails of the broadened
transition spectrum where the rf excitation can have a resonance
offset $\Delta\nu_i\gtrsim10$~kHz. It is often assumed implicitly
that this `hard pulse condition' is fulfilled.

However, while this is readily achievable for small pulse numbers (for a single pulse we require $T_\mathrm{Rabi}\lesssim2/\Delta\nu_\mathrm{inh}\approx25\;\mu$s), the hard pulse condition is increasingly difficult to meet for longer sequences. We consider the Bloch equations of motion in the rotating frame of a static magnetic field $B_\mathrm{z}$
\begin{equation}
\partial \vec{M}(t)/\partial t=\vec{\Omega}\times\vec{M}-\vec{\Gamma}\cdot(\vec{M}-\vec{M}_0)\;\mathrm{,}\label{eq:BEq}
\end{equation}
describing the evolution of magnetisation $\vec{M}=\sum_i\gamma
I_i$ under an angular velocity vector
$\vec{\Omega}=(\Omega_\mathrm{rf}\cos{\varphi},\Omega_\mathrm{rf}\sin{\varphi},2\pi\Delta\nu_i)^\mathrm{T}$
and with relaxation rate
$\vec{\Gamma}=(T_2^{-1},T_2^{-1},T_1^{-1})^\mathrm{T}$ and
equilibrium magnetisation $\vec{M}_0$. Here,
$\Omega_\mathrm{rf}=2\pi/T_\mathrm{Rabi}$ is the resonant Rabi
frequency. We see that even a small resonance offset $\Delta\nu_i$
results in a slight tilt of $\vec{\Omega}$ towards the
$\hat{e}_\mathrm{z}$ axis. As a consequence, an rf pulse of
duration $T_\mathrm{Rabi}$ and carrier phase $\varphi$ will no
longer result in a perfect $2\pi$-rotation of the magnetisation
$\vec{M}$ about $\vec{\Omega}$. Instead, a small rotation angle
error is introduced, which can rapidly lead to non-negligible
effects as the errors of subsequent pulses add up. In practice,
this is reflected in a loss of NMR signal amplitude after a pulse
sequence even in the limit of short free evolution as the
contribution of spins with large $\Delta\nu_i$ to the measured
ensemble magnetisation is reduced.

We consider this effect numerically for the $^{71}$Ga and
$^{75}$As nuclear spin bath ensembles studied in this work. To
quantify the `hardness' (frequency bandwidth) over which a given
pulse sequence is stable against resonance offsets, we evaluate
the evolution of a magnetisation vector $\vec{M}$ under the
sequence including initialisation and readout $\pi/2$-pulses as a
function of $\Delta\nu$. In order to keep our results as general
as possible, we rewrite the resonance offset in terms of the
inverse resonant Rabi period as $\delta = \Delta\nu
T_\mathrm{Rabi}$. We can then describe the rescaled rotation axis
as
$\hat{e}_\Omega=(1+\delta^2)^{-1/2}(\cos{\varphi},\sin{\varphi},-\delta)^\mathrm{T}$.
For added simplicity, our model does not consider any spin
relaxation or dephasing (e.g. through dipolar interaction) and we
set $T_1=T_2=\infty$.

\begin{figure}[htb!]
    \includegraphics[width=0.95\linewidth]{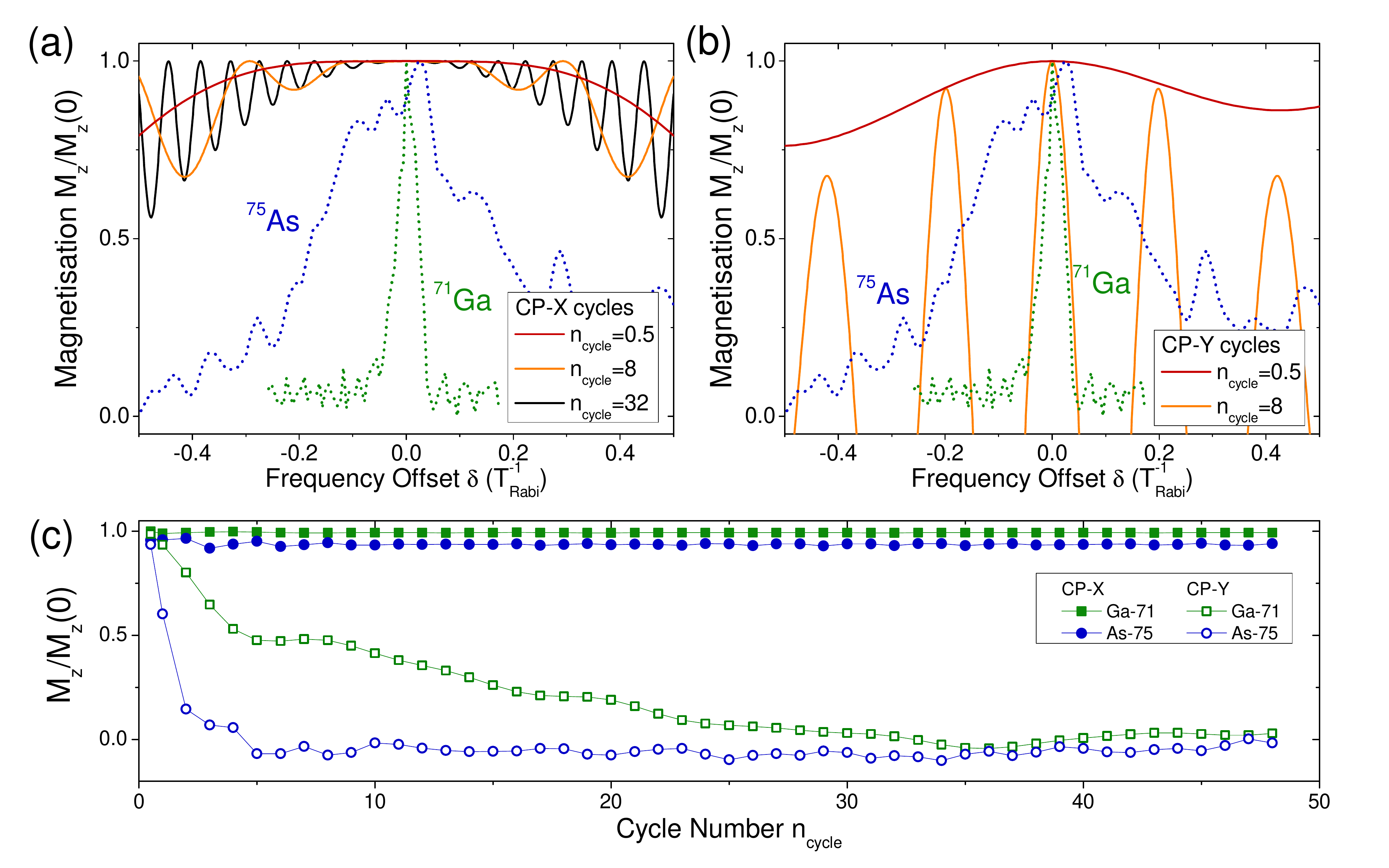}
    \caption{\label{SFig3} (a,b) Normalised magnetisation along the $\hat{e}_\mathrm{z}$ axis after evolution under a series of non-resonant rf pulses as a function of the resonance frequency offset $\delta$ for (a) CP-X and (b) CP-Y sequences with alternating carrier phase. Solid curves show simulated data for different cycle numbers $n_\mathrm{cycle}$. Dashed lines show normalised NMR spectra of the $^{75}$As (blue) and $^{71}$Ga (green) central transitions with frequency axes rescaled by $T_\mathrm{Rabi}$. (c) Weighted average of the normalised magnetisation over the $^{75}$As (circles) and $^{71}$Ga (squares) central transition after a CP-X (solid symbols) or CP-Y (empty symbols) sequence as a function of refocusing cycle number $n_\mathrm{cycle}$.}
\end{figure}

Supplemental Figs~\hyperref[SFig3]{3(a,b)} show the simulated
dependence of the CP-X/Y sequences with alternating pulse carrier
phase discussed in the main text on the frequency offset $\delta$.
Solid lines correspond to sets of simulations with different
$\pi$-pulse numbers. Thus for example, the orange line in
Supplemental Fig.~\hyperref[SFig3]{3(a)} shows the relative change
of the z-component $M_\mathrm{z}/M_\mathrm{z}(0)$ of a
magnetisation vector $\vec{M}$ (with $\vec{M}(0) =
(0,0,1)^\mathrm{T}$) after evolution under a CP-X sequence
($\pi_\mathrm{x}/2-(\pi_\mathrm{x}-\pi_\mathrm{-x})^8-\pi_\mathrm{x}/2$)
for resonance frequency offsets
$-1/2T_\mathrm{Rabi}\leq\delta\leq1/2T_\mathrm{Rabi}$. For
comparison, experimental cw NMR spectra of the $^{71}$Ga and
$^{75}$As central spin transitions rescaled by the respective
shortest Rabi period we could reach experimentally
($T_\mathrm{Rabi}(^{71}\mathrm{Ga})=5.6\;\mu\mathrm{s}$ and
$T_\mathrm{Rabi}(^{75}\mathrm{As})=9.8\;\mu\mathrm{s}$ at rf power
$P=200$~W) are shown as dotted lines.

\begin{figure}[b!]
    \includegraphics[width=0.95\linewidth]{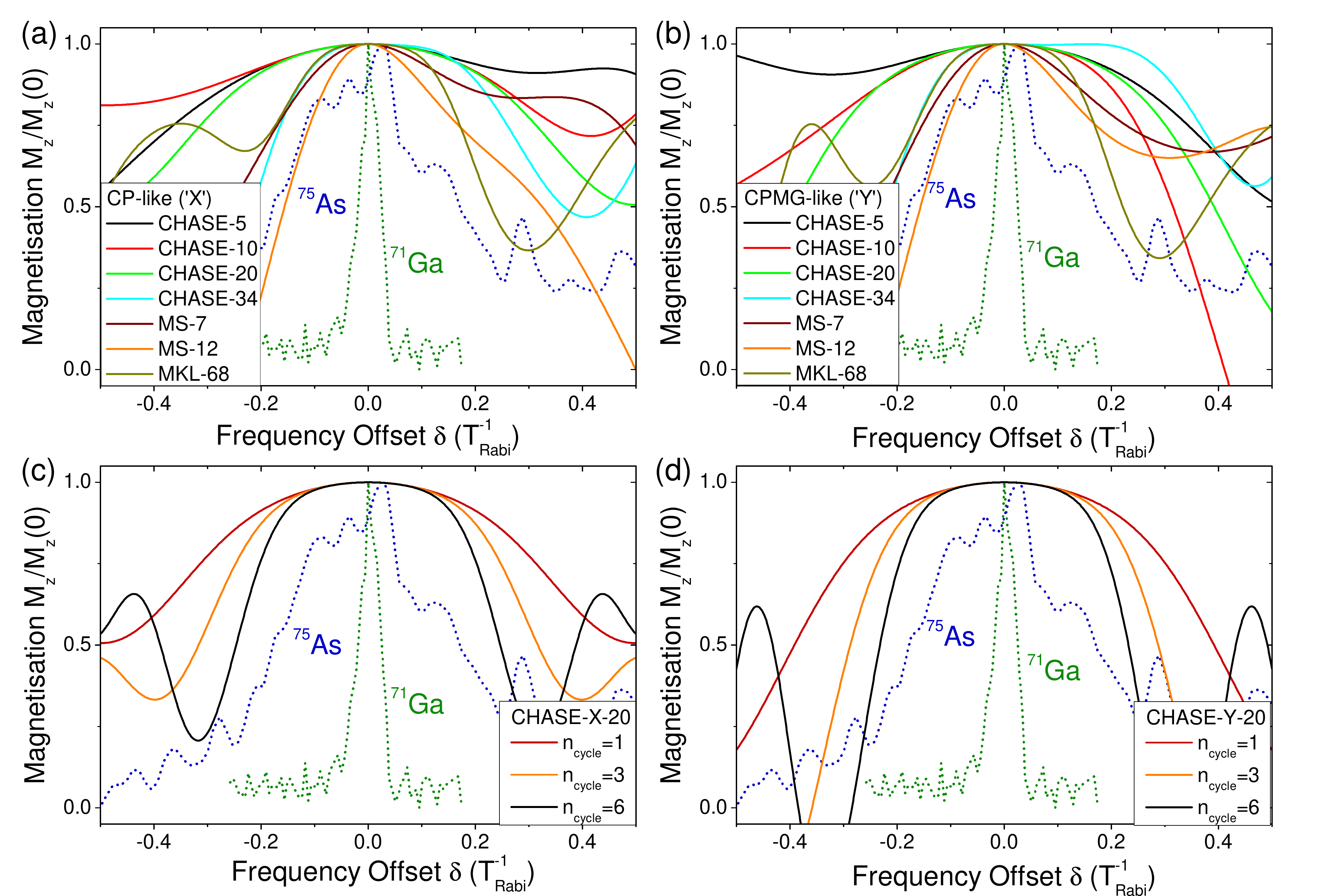}
    \caption{\label{SFig4} Normalised magnetisation along the
        $\hat{e}_\mathrm{z}$ axis after evolution under a series of
        non-resonant rf pulses as a function of the resonance frequency
        offset $\delta$ for (a) CHASE-X and (b) CHASE-Y sequences and
        additional cycles from literature. Dashed lines show normalised
        NMR spectra of the $^{75}$As (blue) and $^{71}$Ga (green) central
        transitions with frequency axes rescaled by $T_\mathrm{Rabi}$.
        (c), (d) Simulated data for CHASE-X/Y-20 with different cycle
        numbers $n_\mathrm{cycle}$.}
\end{figure}

The desired performance is characterized by
$M_\mathrm{z}/M_\mathrm{z}(0)\approx1$ over a wide range of
offsets $\delta$. In this respect, we note that the CP-X cycle
(Supplemental Fig.~\hyperref[SFig3]{3(a)}) is very robust against
frequency offsets: even at large cycle numbers, the magnetisation
vector is largely restored over a large range of offsets. In
practice, this means that we can expect a stable NMR echo
amplitude at short evolution times independent of the number of
applied $\pi$-pulses (as long as additional pulse calibration
errors are negligible). This expected behaviour is shown in
Supplemental Fig.~\hyperref[SFig3]{3(c)}: here, we calculate the
expected experimental echo amplitude from a weighted average of
the normalised final magnetisation over the respective $^{71}$Ga
and $^{75}$As NMR spectra for different cycle numbers
$n_\mathrm{cycle}$. In agreement with the experimental data shown
in Figs 2(e,f) of the main text, the CP-X signal amplitude
(solid symbols) is stable. By contrast, we note in Supplemental
Figs~\hyperref[SFig3]{3(b,c)} that the `bandwidth' within which
the CP-Y sequence can restore the initial magnetisation rapidly
narrows with increasing cycle number $n_\mathrm{cycle}$. Again,
this is in agreement with the experimental observation in the main
text, where the CP-Y NMR signal amplitude $\Delta
E_\mathrm{hf}(\tau_\mathrm{evol}=0)$ rapidly decreases with
increasing $\pi$-pulse number. Although periodic revivals of the
`restored' magnetisation are seen in the orange line of
Supplemental Fig.~\hyperref[SFig3]{3(b)}, these are experimentally
compensated by the contributions of spins at intermediate offsets
$\delta$ where the final magnetisation vector is effectively
flipped. This oscillating behaviour arises as the total pulse
rotation error adds up to multiple precessions of the
magnetisation vector $\vec{M}$ about $\hat{e}_\Omega$.

We run the same simulations for the CHASE sequences introduced in
the main text and for the alternative sequences from literature
analysed in Supplemental Note~\hyperref[sec:sn1]{1}. The results
for the various cycles under `X'- and `Y'-initialisation pulses
are depicted in Supplemental
Figs~\hyperref[SFig4]{4(a)} and~\hyperref[SFig4]{4(b)},
respectively. We note that most sequences have a similar `hard
pulse' bandwidth under both initialisation pulse conditions.
Overall, all of the sequences presented are more stable against
resonance offsets than the CP-Y sequence studied in the main text.
However, the broadband performance of the CP-X sequence
(Supplemental Fig.~\hyperref[SFig3]{3(a)}) remains higher than
that of any CHASE cycle. This is in qualitative agreement with our
experimental results (compare Figs 2(e,f) of the main text).

Additionally, we see that multiple cycles of CHASE-X/Y-20 reduce
the offset tolerance to some extent (Supplemental
Figs~\hyperref[SFig4]{4(c,d)}). This is
confirmed experimentally, as the echo amplitude of the spectrally
broader $^{75}$As ensemble is noticeably reduced with increasing
cycle number in Fig. 2(e) of the main text.

In summary, the reduced NMR signal amplitudes observed in the
experiments on the $^{71}$Ga and $^{75}$As CTs presented in the
main text can be reproduced qualitatively using a simple Bloch
model. We conclude that this is not a fundamental limitation of
the various pulse cycles we studied, but can be attributed to
`soft' rf pulses which for a given spectral broadening of the spin
bath can in principle be avoided by using higher rf excitation
powers. Alternatively, more advanced NMR techniques such as
composite pulses could be implemented in future experiments to
increase the `hardness' of the applied
pulses\cite{levitt1979s,tycko1983s}.

\clearpage

\section{\label{sec:sn4}Numerical simulation of the nuclear spin evolution under pulsed radiofrequency manipulation}

In this Note we describe the details of the numerical simulation of the quantum mechanical evolution dynamics of the interacting nuclear spin bath.

\subsection{\label{sec:sn4-1}The model.}

We consider once more the model introduced in Supplemental
Note~\hyperref[sec:sn1]{1}, where the evolution of the
wavefunction $\psi(t)$ describing the state of the nuclear spin
bath in the rotating frame of an external magnetic field
$B_\mathrm{z}$ is determined by:
\begin{equation}
\begin{split}
\partial\psi(t)/\partial t=-(\textrm{i}/\hbar)\mathcal{H}(t)\psi(t)\;\mathrm{,}\\
\mathcal{H}(t) = \mathcal{H}_\textrm{0}^\mathrm{z} +\mathcal{H}_\textrm{d}^\mathrm{zz} + \mathcal{H}_\textrm{rf}(t)\;\mathrm{.}\label{eq:SU}
\end{split}
\end{equation}
As before, the Hamiltonian $\mathcal{H}(t)$ is composed of a term
$\mathcal{H}_\textrm{0}^\mathrm{z}$ describing here quadrupolar
interaction with electric field gradients, a nuclear dipolar
coupling term $\mathcal{H}_\textrm{d}^\mathrm{zz}$ and a
radio-frequency (rf) term $\mathcal{H}_\textrm{rf}(t)$.

We consider half-integer spins $I$ and simulate evolution only of
the $I_\mathrm{z}=\pm1/2$ subspace corresponding to the NMR
experiments on the central transition. The effect of the
quadrupolar interaction (more specifically of its second order
term) on the $I_\mathrm{z}=\pm1/2$ manifold is equivalent to an
additional magnetic field that changes the Larmor frequency of the
$i$-th nucleus by $\Delta\nu_i$. The quadrupolar term can then be
written explicitly as:
\begin{equation}
\mathcal{H}_\textrm{0}^\mathrm{z} = 2\pi\hbar\sum_{i=1}^{N}\Delta\nu_i I_{\mathrm{z},i}\;\mathrm{,}\label{eq:HQ}
\end{equation}
where the summation goes over all $N$ nuclei.

We consider the case of high magnetic field (significantly larger than the local dipolar field), so that the nuclear-nuclear interaction is described by the truncated dipole-dipole Hamiltonian:
\begin{equation}
\mathcal{H}_\textrm{d}^\mathrm{zz}=\frac{\mu_0}{4\pi}\hbar^2\gamma^2\sum_{i<j}{\frac{x_{i,j}^2+y_{i,j}^2-2z_{i,j}^2}{(x_{i,j}^2+y_{i,j}^2+z_{i,j}^2)^{5/2}}\left[I_{\mathrm{z},i}I_{\mathrm{z},j}-\frac{(I+1/2)^2}{2}(I_{\mathrm{x},i}I_{\mathrm{x},j}+I_{\mathrm{y},i}I_{\mathrm{y},j})\right]}\;\mathrm{,}\label{eq:Hdd}
\end{equation}
where $\mu_0=4\pi\times10^{-7}$~N/A$^{2}$ is the magnetic
constant, $\gamma$ is the nuclear gyromagnetic ratio,
$(x_{i,j},y_{i,j},z_{i,j})$ is the vector connecting spins $i$ and
$j$ and the summation goes over all pairs of non-identical nuclei.
We ignore here any possible contributions from pseudo-dipolar or
exchange interactions.

The effect of the rf field is described by:
\begin{equation}
\mathcal{H}_\textrm{rf}(t)=2\pi\hbar(I+1/2)\sum_{i=1}^{N}{(\nu_{1,\mathrm{x}}(t)    I_{\mathrm{x},i}+\nu_{1,\mathrm{y}}(t)I_{\mathrm{y},i})}\;\mathrm{,}\label{eq:Hrf}
\end{equation}
where parameters $\nu_{1,\mathrm{x}}(t)$ and $\nu_{1,\mathrm{y}}(t)$ characterise the amplitude of the rf magnetic field along the $x$ and $y$ axes of the rotating frame respectively. The parameters $\nu_{1,\mathrm{x}}(t)$ and $\nu_{1,\mathrm{y}}(t)$ are piecewise functions of time describing the variation of the rf field amplitude during the pulse sequence.

In the above equations~\ref{eq:HQ}, \ref{eq:Hdd} and~\ref{eq:Hrf}
the operators $I_\mathrm{x}$, $I_\mathrm{y}$ and $I_\mathrm{z}$
are spin-1/2 Pauli matrices acting on the effective spin-1/2
subspace of the spin states with spin projections
$I_\mathrm{z}=\pm1/2$. The factors $(I+1/2)$ in
equations~\ref{eq:Hdd} and~\ref{eq:Hrf} (which differ them from
equations \ref{SEq2}, \ref{SEq3}, \ref{SEq4}) arise from the matrix elements of the
$I_\mathrm{x}$ and $I_\mathrm{y}$ operators of the full spin-$I$
nuclei that are projected on to the $I_\mathrm{z}=\pm1/2$
subspace. All quantities are in SI units, e.g.\ frequencies $\nu$
are in hertz and Hamiltonians are in joules.

The evolution of the nuclear spin magnetisation is calculated by
direct propagation of the Schr\"{o}dinger equation
(Eq.~\ref{eq:SU}) using the 6-th order Runge-Kutta method
implemented in Mathematica 10.3 and 11.0.1. The Hamiltonian
matrices are stored and handled as `sparse arrays' for improved
computation efficiency. For a given system of few spins, direct
evaluation of Eq.~\ref{eq:SU} gives an exact evolution of the
wavefunction, thus yielding a complete description of the nuclear
spin dynamics. The maximum number of spins $N$ that can be
simulated with this direct approach is severely limited by the
computation resources, since the memory and the computation time
scale approximately as $\sim N^2$ and $\sim N^4$ respectively.
Thus when performing simulations on a `toy' system with small $N$,
the choice of the parameters and initial states becomes important
for obtaining results that are relevant for systems with much
larger $N$ (such as quantum dots with $N\geq10000$). The choice of
parameters is discussed in the following section.

\subsection{\label{sec:sn4-2}Model parameters for simulation of the nuclear spin dynamics in quantum dots.}

For simulations, nuclear spins are placed at the nodes of the
face-centered cubic (fcc) lattice. One nucleus is placed at the
origin $x=y=z=0$ and the other nuclei are selected from its
nearest neighbors -- this way the nuclear spin cluster is kept as
`spherical' and `dense' as possible which allows approximating the
complexity and the magnitude of the dipolar interaction in a 3D
lattice. Example clusters with $N=6$, $N=12$ and $N=19$ spins are
shown in Supplemental Fig.~\hyperref[Fig:TestData]{5(a)}. The
nuclei are taken to be $^{75}$As with gyromagnetic ratio
$\gamma=2\pi\times7.29\times10^6$~rad/s. The lattice constant of
the fcc lattice is taken to be $a_0=0.564786$~nm, corresponding to
GaAs at a temperature of~$\sim$4~K.

Inhomogeneous quadrupolar interaction is introduced by varying the
Larmor frequency $\nu_\textrm{L}$ of each spin (Eq.~\ref{eq:HQ}).
Firstly, the Larmor frequency of the $i$-th spin $i\in[1,N]$ in
the rotating frame is set to be
$\Delta\nu_i=\nu_\textrm{Q}(i-1/2-N/2)$. This equidistant set of
$\Delta\nu_i$ is then rescaled as $\Delta\nu_i\rightarrow
k_1\times\Delta\nu_i$, preserving the mean Larmor frequency
$\overline{\Delta\nu_i}=0$. The factor $k_1$ is chosen as an
implementation of a uniform random distribution in the range
$[0.77..1.3]$. At the next step, each frequency $\Delta\nu_i$ is
modified further by adding a random offset
$k_{2,i}\times\nu_\textrm{Q}$ with $k_{2,i}$ selected randomly for
each nucleus from a uniform distribution in the range
$[-0.325,0.325]$. The purpose of such a randomisation using
parameters $k_1$ and $k_{2,i}$ is to eliminate spurious periodic
`beatings' in the nuclear spin dynamics arising from the small
number of the nuclear spins $N<20$ used in the simulations. Such
`beatings' are not present in experimental decay curves on real
III-V quantum dots where $N>10000$, but similar features are
observed for defect spins in dilute nuclear spin baths (e.g.\ NV
centres in diamond), where electron-nuclear hyperfine coupling
leads to periodic coherence collapses\cite{mims1972s, jelezko2004s}.
The physical meaning of the randomisation procedure can be seen as
follows: The quantum dot can be viewed as built of a large number
of clusters, each containing $N$ spins with a different random
distribution of quadrupolar frequencies. The experimentally
measured NMR signal is an average over all such clusters, which is
simulated by Monte-Carlo averaging over $k_1$ and $k_{2,i}$ in the
numerical calculations. The step $\nu_\textrm{Q}$ for the
equidistant spacing of Larmor frequencies is chosen to be large
compared to the dipolar coupling $\nu_{ij}$ of any two spins, so
that the suppression of dipolar flip-flops arising from
quadrupolar interaction (characteristic of self-assembled quantum
dots \cite{chekhovich2015s}) can be efficiently simulated. We
typically use $\nu_\textrm{Q}=2000$~Hz, which is chosen
empirically by observing that no change in spin dynamics occurs
when $\nu_\textrm{Q}$ is increased further. The ranges for $k_1$
and $k_{2,i}$ are also chosen from trial simulations to be large
enough to suppress spurious `beatings' while still small enough to
ensure that the minimum difference between any $\Delta\nu_i$ is
large enough to emulate strongly inhomogeneous quadrupolar
interaction. When simulating spin dynamics of the nuclei in the
absence of quadrupolar effects we use the above procedure with
$\nu_\textrm{Q}$, $k_1$ and $k_{2,i}$ set to zero.

For initialisation of the nuclear spin system we use the following
procedure. Each of the $N$ nuclear spins is randomly initialised
in one of the four single-spin eigen states with
$I_\mathrm{z}=-3/2..+3/2$. The probabilities to find each nucleus
in $I_\mathrm{z}=\pm1/2$ states and $I_\mathrm{z}=\pm3/2$ are
taken to be 60\% and 40\% respectively. The probabilities for the
$I_\mathrm{z}=+1/2$ and $I_\mathrm{z}=-1/2$ states are taken to
produce 75\% polarisation degree in the $I_\mathrm{z}=\pm1/2$
subensemble. Such a choice of probabilities corresponds closely to
the experimental conditions where optical pumping inducing nuclear
spin polarisation degree of $\sim$50\% is followed by adiabatic
radiofrequency sweeps exchanging the populations of
$I_\mathrm{z}=\pm1/2$ and $I_\mathrm{z}=\pm3/2$ states
\cite{chekhovich2015s}. The nuclei in the $I_\mathrm{z}=\pm3/2$
states are then ignored when simulating the spin dynamics of the
$I_\mathrm{z}=\pm1/2$ states. This is justified since the
$I_\mathrm{z}=\pm3/2$ states have very long correlation times
($\tau_\textrm{c}\sim$10~s, Ref.~\cite{waeber2016s}) and act on the
$I_\mathrm{z}=\pm1/2$ spins simply as a source of quasistatic
local magnetic fields which are already taken into account by the
inhomogeneous spread of the Larmor frequencies $\Delta\nu_i$. This
initialisation procedure gives a tensor product random state which
is not an eigenstate but where each nucleus is in a single-spin
eigenstate with $I_\mathrm{z}=+1/2$ or $I_\mathrm{z}=+1/2$. In
each simulation run (Monte-Carlo sample) the initial function is
constructed by repeating the above procedure and creating a linear
superposition of 1000 basic random states with random complex
weighting coefficients. Such a highly entangled pure state with
finite polarisation along $z$ direction has `self-averaging'
properties arising from `quantum parallelism'\cite{schliemann2002s}
and allows for faster convergence of the Monte-Carlo simulations.

The typical number of Monte-Carlo samples is 1000. For each
Monte-Carlo sample a set of nuclear frequency shifts $\Delta\nu_i$
is generated (with random parameters $k_1$ and $k_{2,i}$), the
wavefunction is then initialised into a random superposition state
as described above. The time evolution of the wavefunction is then
calculated numerically from the Schr\"{o}dinger
equation~\ref{eq:SU} with a Hamiltonian whose time dependence is a
piecewise function determined by the rf pulse sequences. The
overall time dependence of the nuclear spin polarisation is
calculated by averaging over the Monte-Carlo samples. In all
simulations the nuclei are first initialised in a state polarised
along the $\hat{e}_\mathrm{z}$ axis, then a single $\pi/2$-pulse
is used to rotate the polarisation into the $xy$ plane, then a
time sequence consisting of rf pulse rotations and free evolution
periods is simulated, finally a single $-\pi/2$-pulse is used to
rotate the magnetisation. All of the studied NMR pulse sequences
are cyclic, i.e.\ in the limit of a short free evolution and ideal
rf pulses the magnetisation is returned into its original state
along the $\hat{e}_\mathrm{z}$ axis (or into a state with inverted
$z$-magnetisation for HE-Y, the Meiboom-Gill version of Hahn
echo). In simulations we use both ideal (infinitely short, or
`hard') and non-ideal (finite duration, or `soft') rectangular rf
pulses. The total free evolution time $\tau_\mathrm{evol}$ is
varied, and for each value of $\tau_\mathrm{evol}$ the final value
of the nuclear magnetisation $\langle I_\mathrm{z} \rangle$ along
the $\hat{e}_\mathrm{z}$ axis is computed. The resulting time
dependence $\langle I_\mathrm{z} (\tau_\mathrm{evol})\rangle$
reflects the process of nuclear spin decoherence and can be used
to derive the coherence time $T_2$.

\clearpage

\subsection{\label{sec:sn4-3}Examples of simulations of the nuclear spin dynamics in quantum dots: dependence on the number of spins $N$.}

\begin{figure*}
    \includegraphics[width=1.0\linewidth]{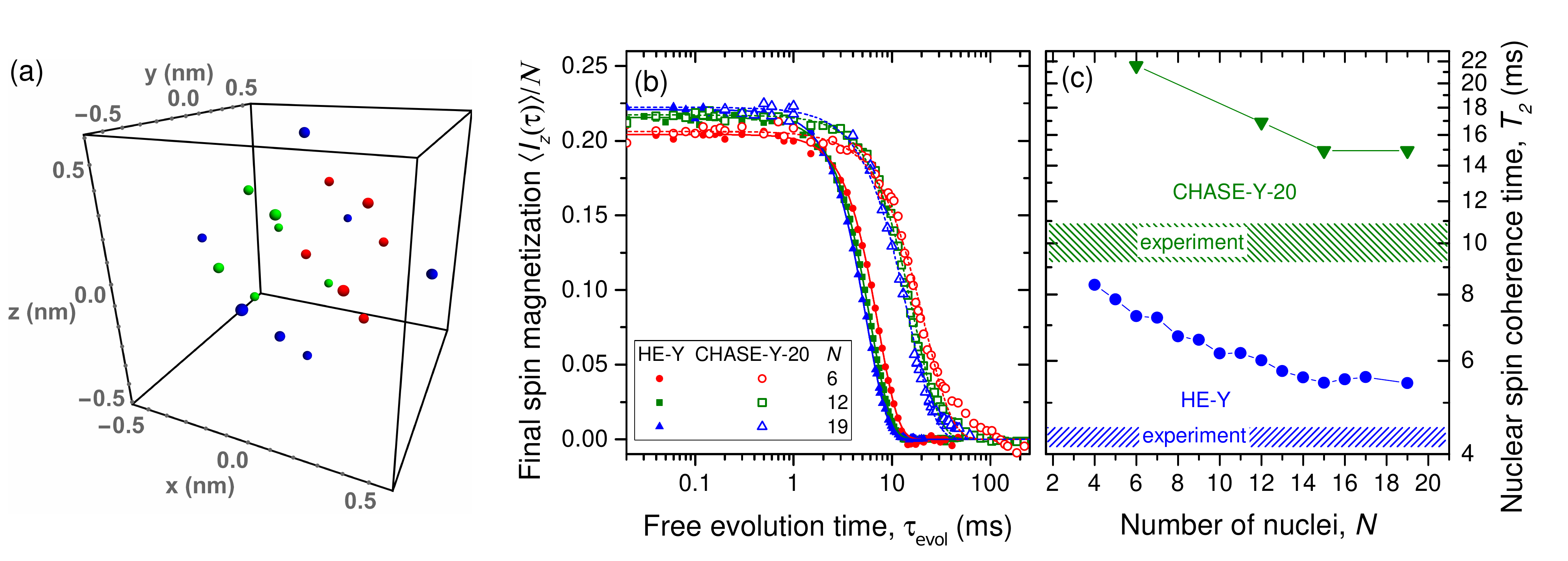}
    \caption{(a) Spatial geometry of the $^{75}$As nuclear spin cluster used for numerical simulations. Red balls show a configuration with $N$=6 spins, red and green with $N$=12 spins, red, green and blue combined together form a cluster with $N$=19 spins. (b) Simulated dependence of the final nuclear spin magnetisation $\langle I_\mathrm{z} (\tau_\mathrm{evol})\rangle$ (normalised by
        the number of spins $N$) on the total free evolution time
        $\tau_\mathrm{evol}$ under Hahn Echo (HE-Y, solid symbols) and
        CHASE-Y-20 pulse sequences computed for $N$=6 spins (circles),
        $N$=12 spins (squares), and $N$=19 spins (triangles). A
        Meiboom-Gill version of Hahn echo sequence
        ($\pi/2_\mathrm{y}-\tau_\mathrm{evol}/2-\pi_\mathrm{x}-\tau_\mathrm{evol}/2-\pi/2_\mathrm{-y}$) is used with a
        $\pi/2$-shift between the phases of the $\pi/2$- and $\pi$-pulses.
        Lines show the fitting used to derive nuclear spin decoherence
        times $T_2$. (c) Nuclear spin decoherence times $T_2$ derived from
        numerical simulations plotted as a function of the number of
        nuclear spins for Hahn Echo (circles) and CHASE-Y-20 (triangles)
        pulse sequences. Shaded areas show experimentally measured
        decoherence times of $^{75}$As spins in a self-assembled quantum
        dot.}
    \label{Fig:TestData}
\end{figure*}

We now give several examples of the results obtained from the
above described numerical simulation procedure. Several simulated
$\langle I_\mathrm{z}(\tau_\mathrm{evol}) \rangle$ curves are
shown in Supplemental Fig.~\hyperref[Fig:TestData]{5(b)} for a
Meiboom-Gill version Hahn echo (HE-Y, solid symbols) and
CHASE-Y-20 (open symbols) pulse sequences -- these are computed
for clusters with different numbers of nuclei shown in
Supplemental Fig.~\hyperref[Fig:TestData]{5(a)} for the case of
large inhomogeneous quadrupolar interaction
($\Delta\nu_i\gg\nu_{ij}$). Lines show fitting using compressed
exponential functions. For the Hahn echo sequence the decay is
close to Gaussian (characterised by compression factor
$\beta\approx2.0-2.1$), while for CHASE-Y-20 the best fit is for
$\beta\approx1.56-1.67$, and some deviation from a
mono-exponential decay is observed, especially at small $N$.

The nuclear spin decoherence times $T_2$ derived from the fits as
in Supplemental Fig.~\hyperref[Fig:TestData]{5(b)} are shown in
Supplemental Fig.~\hyperref[Fig:TestData]{5(c)} by the symbols and
are compared to the experimental values for $^{75}$As spins in
self-assembled quantum dots (shaded areas). It can be seen that
the number of spins affects the overall timescale of the nuclear
spin decoherence -- for larger $N$ the nuclear spin decoherence is
faster as the interaction with a larger number of neighbors is
taken into account. However, the overall trend in variation of
$T_2$ under different pulse sequences is found to be robust
against $N$. For example, while the decoherence times $T_2$ depend
on $N$ as shown in Fig.~\hyperref[Fig:TestData]{5(c)}, the ratio
of the $T_2$ values under CHASE-Y-20 and Hahn echo sequences is
nearly independent of $N$, ranging between $\sim2.74$ for $N=19$
and $\sim3.0$ for $N=6$ which is in good agreement with the
experimental ratio of $\sim2.8$. These test results justify the
use of relatively small $N$ -- for most simulations in this work
we employ $N=12$ giving a good compromise between accuracy and
computation time. The fact that the $T_2$ simulated for $N=12$
differs from the experimental $T_2$ on a system with $N\sim10000$
only by $\sim$50\% indicates the robustness of our approach. Thus
our simulations (i) give a good quantitative numerical estimate of
the absolute $T_2$ values, and (ii) provide an excellent tool for
examining the effect of various pulse sequences on $T_2$.

\subsection{\label{sec:sn4-4}Procedure for derivation of the nuclear spin coherence times and echo amplitudes from the results of numerical simulations.}

\begin{figure}
    \includegraphics[width=1.0\linewidth]{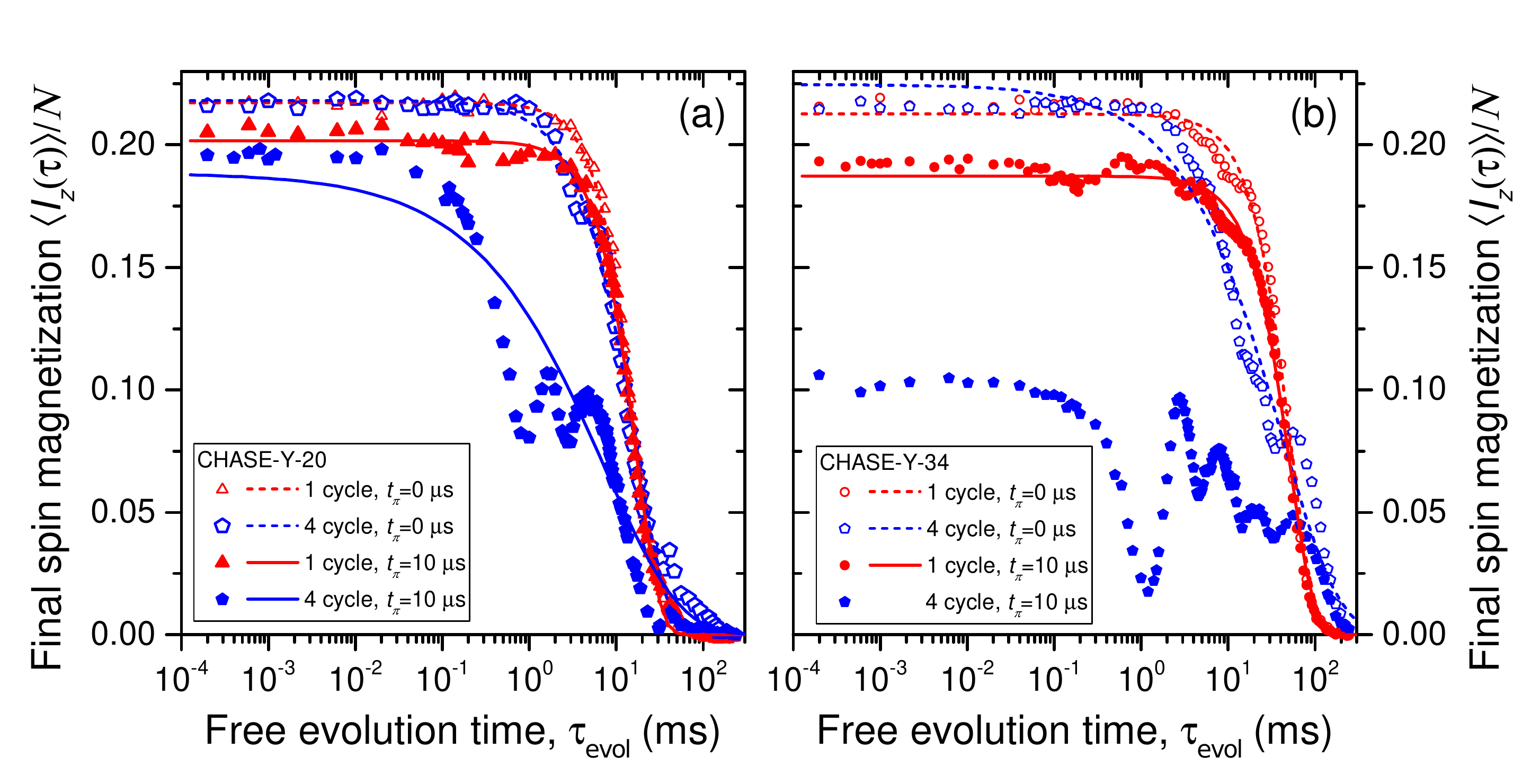}
    \caption{Simulated dependence of the final nuclear spin
        magnetisation $\langle I_\mathrm{z} (\tau_\mathrm{evol})\rangle$
        (normalised by the number of spins $N$) on the total free
        evolution time $\tau_\mathrm{evol}$ under CHASE-Y-20 (a) and
        CHASE-Y-34 (b) pulse sequences computed for $N=$12 spins. The
        results are presented for infinitely short ($t_\pi=0$, open
        symbols) and finite ($t_\pi=10~\mu$s, solid symbols) control
        pulses. The calculations were performed for 1 cycle (triangles)
        and for 4 cycles (pentagons) of the sequence. Lines show best
        least-squares fits using compressed exponents. In the case of 4
        cycles of CHASE-Y-34 with $t_\pi=10~\mu$s the imperfect pulse
        rotations result in significant loss of transverse nuclear spin
        magnetisation even at short $\tau_\mathrm{evol}$ -- this prohibits
        unambiguous definition of the coherence time $T_2$, thus no
        fitting results are shown.} \label{Fig:ChaseData}
\end{figure}

We now present the raw data of the numerical simulations for the
CHASE sequences (Supplemental Fig.~\ref{Fig:ChaseData}) and
discuss the procedure for analysing the raw data and deriving the
spin bath coherence times $T_2$ and the echo amplitudes $\langle
I_\mathrm{z}(\tau_\mathrm{evol}=0) \rangle$ in the limit of short
free evolution time $\tau_\mathrm{evol}\rightarrow0$. Supplemental
Fig.~\hyperref[Fig:ChaseData]{6(a)} shows the simulated spin bath
dynamics under CHASE-Y-20. The results are presented for ideal
infinitely short (`hard') rf control pulses (open symbols) and for
the finite (`soft') rectangular pulses (solid symbols, $\pi$-pulse
length of $t_\pi=10~\mu$s). The calculations were performed for 1
cycle of the sequence (triangles) and for 4 cycles (pentagons).
Lines show best least-squares fits using compressed exponents.
These fits are used to derive the spin bath coherence times $T_2$.
While for 1 cycle the fit is good, for more complex conditions,
e.g. 4 cycles and $t_\pi=10~\mu$s, there is a considerable
deviation between numerical experiment and exponential fits -- in
such cases the $T_2$ times are still derived from fitting but
should be treated as approximate values.

Supplemental Fig.~\hyperref[Fig:ChaseData]{6(b)} shows further
results for the spin bath dynamics under the CHASE-Y-34 pulse
sequence. Here deviation from the exponential fit is observed for
4 cycles even at $t_\pi=0$ while at $t_\pi=10~\mu$s the
oscillations and reduction of the echo amplitude at short free
evolution time $\tau_\mathrm{evol}\rightarrow0$ are particularly
pronounced. This requires care when deriving decoherence
parameters. Firstly, in our analysis the echo amplitude $\langle
I_\mathrm{z}(\tau_\mathrm{evol}=0)\rangle$ is derived not from
fitting but rather by taking the average spin magnetisation
$\langle I_\mathrm{z} \rangle$ (normalised by the number of nuclei
$N$) at short free evolution times $\tau_\mathrm{evol}<5~\mu$s --
this definition of $\langle
I_\mathrm{z}(\tau_\mathrm{evol}=0)\rangle$ is not affected by
deviation of the spin decay from the exponential model. Secondly,
for any cyclic pulse sequences with ideal `hard' pulses
($t_\pi=0$) the resulting magnetisation $\langle
I_\mathrm{z}(\tau_\mathrm{evol}=0)\rangle$ after the sequence with
short free evolution $\tau_\mathrm{evol}\rightarrow0$ is by
definition equal to the initial magnetisation $\langle
I_\mathrm{z}(t=0)\rangle$ before the pulse sequence is applied (in
the studied example $\langle
I_\mathrm{z}(t=0)\rangle/N\approx0.217$), while for non-ideal
pulses ($t_\pi>0$) nuclear spin magnetisation at
$\tau_\mathrm{evol}\rightarrow0$ may be lost simply due to the
imperfect spin rotations (i.e. due to the `soft' pulse
conditions). Such imperfect rotations mean that the spin bath
states during free evolution periods of finite duration
$\tau_\mathrm{evol}>0$ deviate from the desired sequence. Under
such conditions (e.g.\ $t_\pi=10~\mu$s in Supplemental
Fig.~\ref{Fig:ChaseData}) the reduction in $\langle
I_\mathrm{z}(\tau_\mathrm{evol})\rangle$ is not related to
decoherence as such, prohibiting any unambiguous definition for
$T_2$.

Taking into account the above arguments we establish an approach
to the analysis of the numerical results which can be summarised
as follows: The echo amplitude $\langle
I_\mathrm{z}(\tau_\mathrm{evol}=0)\rangle$ is derived by averaging
the $I_\mathrm{z}$ over short free evolution times
$\tau_\mathrm{evol}<5~\mu$s. For echo amplitudes $\langle
I_\mathrm{z}(\tau_\mathrm{evol}=0)\rangle$ below 70\% of the
initial magnetisation $\langle I_\mathrm{z}(t=0)\rangle$ the
coherence time $T_2$ is undefined, while for $\langle
I_\mathrm{z}(\tau_\mathrm{evol}=0)\rangle$ above this threshold,
the $T_2$ is derived from fitting with compressed exponential
functions. Moreover, in the main text and the subsequent
discussion we present echo amplitudes at short free evolution
times $\langle I_\mathrm{z}(\tau_\mathrm{evol}=0)\rangle$
normalised by the initial magnetisation $\langle
I_\mathrm{z}(t=0)\rangle$.

\begin{figure}[htb!]
    \includegraphics[width=0.8\linewidth]{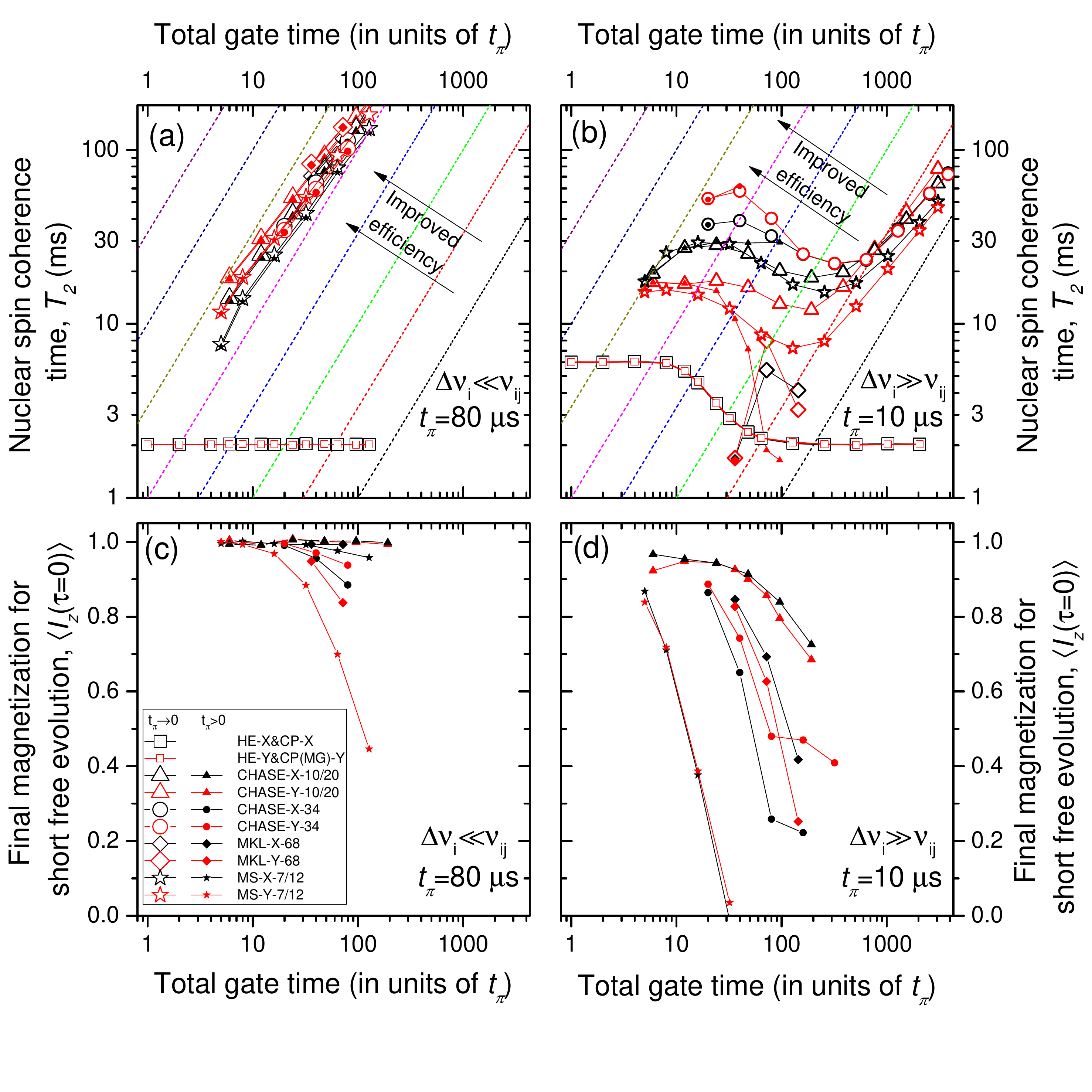}
    \caption{\label{Fig:Literaturesequences} Results of the nuclear spin decoherence numerical simulations for $N=12$ dipolar coupled $^{75}$As nuclear spins under rf pulse control sequences for the case of small $\Delta\nu_i\ll\Delta_{ij}$ (a,c) and large $\Delta\nu_i\gg\Delta_{ij}$ (b,d) quadrupolar broadening. Symbols in figures (a) and (b) show the nuclear spin coherence times $T_2$ for different pulse sequences as a function of the total pulse (gate) time in units of $t_\pi$. The plot for each type of sequence is obtained by varying the number of cycle repeats; for Hahn Echo (HE), MS-7 and CHASE-10 we consider only one cycle and combine the data with CP, MS-12 and CHASE-20 respectively. The dashed lines represent constant efficiencies of the pulse sequences, defined as coherence time to gate time ratio. The gate time dependencies of the final magnetisation (echo amplitude) at short free evolution $\langle I_\mathrm{z}(\tau_\mathrm{evol}=0) \rangle$ are shown in (c) and (d), the values are normalised by the magnitude of the initial magnetisation $\langle I_\mathrm{z}(t=0)\rangle$. Simulations are carried out for both infinitely short ($t_\pi=0$, open symbols) and finite pulses (solid symbols), where we set $t_\pi=80~\mu$s for $\Delta\nu_i\ll\nu_{ij}$ and $t_\pi=10~\mu$s for $\Delta\nu_i\gg\nu_{ij}$. The $\langle I_\mathrm{z}(\tau_\mathrm{evol}=0) \rangle$ values in (c) and (d) are plotted only for $t_\pi>0$ since at $t_\pi=0$ one has $\langle I_\mathrm{z}(\tau_\mathrm{evol}=0)\rangle=1$ for any cyclic pulse sequence by definition.}
\end{figure}

\subsection{\label{sec:sn4-5} Suppression of the nuclear spin fluctuations under various pulse sequences: results of numerical simulations.}

In the main text we present the results of numerical simulations
for the pulse sequences used in the experimental work. Simulations
are in good agreement with the experiment and confirm robust
extension of the nuclear spin coherence time $T_2$ under CHASE
pulse sequences. In this section we present simulated nuclear spin
dynamics under alternative pulse sequences reported in the
literature and compare their performance to CHASE.

Having discussed in the previous section how echo amplitudes
$\langle I_\mathrm{z}(\tau_\mathrm{evol}=0)\rangle$ and coherence
times $T_2$ are derived, we now examine their dependence on the
control pulse sequence parameters. The results of the simulations
are summarised in Supplemental Fig.~\ref{Fig:Literaturesequences}
for the cases of small inhomogeneous quadrupolar interaction
($\Delta\nu_i\ll\nu_{ij}$, panels a, c) and large inhomogeneous
quadrupolar interaction ($\Delta\nu_i\gg\nu_{ij}$, panels b, d).
Three types of sequences are presented: (i) CHASE-10/20 as
proposed in this work, (ii) the 7 and 12 pulse sequences proposed
theoretically by Moiseev and
Skrebnev\cite{moiseev2015s,moiseev2015bs} and labeled MS-7/12 here,
(iii) the sequence consisting of 8 MREV-8 pulse trains interwoven
with four phase-refocusing $\pi$-pulses used by Maurer, Kucsko et
al.\cite{maurer2012s} in experiments on NV centers in diamond and
labeled MKL-68. The results for Hahn Echo (HE) and Carr-Parcell
(CP) sequences are shown as well for a reference. The numbers in
the sequence labels stand for the total number of rf control
pulses in one cycle. All results in Supplemental
Fig.~\ref{Fig:Literaturesequences} are plotted as a function of
the total duration of the control rf pulse sequence (total gate
time) in the units of the $\pi$-pulse duration $t_\pi$. Similar to
the way the results are presented in the main text, we combine
CHASE-10 with CHASE-20 and MS-7 with MS-12: the points with the
shortest total gate time correspond to one cycle of CHASE-10 and
MS-7 sequences, while points with larger gate times correspond to
integer numbers of repeated cycles of CHASE-20 and MS-12. For each
sequence we consider two cases: with nuclear magnetisation
initialised by a $\pi/2$-pulse along the same $\hat{e}_\mathrm{x}$
axis as the $\pi$-pulses of the sequence (`-X' sequences) and with
initialisation along the $\hat{e}_\mathrm{y}$ axis, orthogonal to
that of the $\pi$ pulses (Meiboom-Gill version, labeled `-Y').

\subsubsection{\label{sec:sn4-5-1} The case of zero inhomogeneous resonance broadening.}

We first examine the case of zero inhomogeneous resonance
broadening $\Delta\nu_i\ll\nu_{ij}$ (negligible quadrupolar
effects or chemical shifts) as shown in Supplemental
Fig.~\hyperref[Fig:Literaturesequences]{7(a),(c)}. It follows from
Supplemental Fig.~\hyperref[Fig:Literaturesequences]{7(a)} that
all three types of sequences can be used to achieve arbitrarily
long nuclear spin coherence time: the $T_2$ increases
approximately linearly with the increasing number of sequence
repetitions (increasing total gate time). This is largely expected
from AHT -- when the number of cycles is increased, the    cycle
duration $t_c$ is reduced, and the average Hamiltonian converges
to its remaining zeroth order term
$\bar{\mathcal{H}}_\mathrm{d}^{(0)}$. As shown in Supplemental
Tables~\ref{tab1} and~\ref{tab2}, this term vanishes for $t_\pi=0$
for all studied sequences. For a given total gate time, the $T_2$
values are very close for all three types of sequences for initial
magnetisation along either $x$ or $y$ axes -- the difference is
less than a factor of 2. However, the performance of the sequences
is notably different when non-ideal pulses ($t_\pi>0$) are
considered. For both MKL and MS sequences a pronounced loss of
magnetisation $\langle I_\mathrm{z}(\tau_\mathrm{evol}=0)\rangle$
at short free evolution (echo amplitudes) is observed for the
Meiboom-Gill (`Y') versions of the sequences when the number of
cycles is increased (Supplemental
Fig.~\hyperref[Fig:Literaturesequences]{7(c)}) -- this means that
strong nuclear spin decoherence is induced by the finite `soft'
control pulses irrespective of decoherence during free evolution
between the pulses. By contrast the CHASE sequences show robust
performance for an arbitrary direction of the initial nuclear spin
magnetisation for the total gate times of up to $\sim 200 t_\pi$
studied here, thus demonstrating their capability to dynamically
`freeze' arbitrary fluctuation of the transverse nuclear
magnetisation.

\subsubsection{\label{sec:sn4-5-2} The case of large inhomogeneous resonance broadening.}

The case of large inhomogeneous resonance broadening
$\Delta\nu_i\gg\nu_{ij}$ (e.g.\ strong quadrupolar effects) is
presented in Supplemental
Fig.~\hyperref[Fig:Literaturesequences]{7(b),(d)}. We start by
examining the coherence times under ideal `hard' control pulses
($t_\pi=0$, open symbols in Supplemental
Fig.~\hyperref[Fig:Literaturesequences]{7(b)}). The MKL sequence
exhibits reduced $T_2$ times, which are even shorter (for 1 cycle)
than in the case of simple $\pi$ pulse trains (Carr-Parcell
sequences, CP). This is likely due to the fact that the MKL
sequence was not designed to be applied to strongly inhomogeneous
spin systems in the first place. By contrast, all of the CHASE and
MS sequences provide enhancement in $T_2$ compared to Hahn echo
and CP and show a similar non-monotonic behaviour on the total
gate time which is also presented in Fig. 3(b) of the main text
for CHASE-10/20 sequences. For the total gate times up to $\sim100
t_\pi-200 t_\pi$ the nuclear spin coherence time $T_2$ is seen to
decrease. Such reduction is also observed for the CP sequences and
is interpreted to arise from fast rotations of the spins by the rf
pulses which lead to an effectively shortened spin lifetime and
broadened nuclear spin transitions. Such a broadening can
compensate for the energy mismatch between the spins induced by
the quadrupolar inhomogeneity and restores the dipolar exchange
spin-spin flip-flops. This interpretation is readily confirmed by
examining the CP results: for a large number of pulse cycles (with
the total gate time $\gtrsim100 t_\pi$) the $T_2$ of the
inhomogeneous ($\Delta\nu_i\gg\nu_{ij}$, Supplemental
Fig.~\hyperref[Fig:Literaturesequences]{7(b)}) nuclear spin bath
reduces to exactly the value of $T_2\approx2.02$~ms observed for
the homogeneous bath ($\Delta\nu_i\ll\nu_{ij}$, Supplemental
Fig.~\hyperref[Fig:Literaturesequences]{7(a)}) where dipolar
flip-flops are allowed. When the number of CHASE or MS sequence
cycles is increased further ($\gtrsim500 t_\pi$ in Supplemental
Fig.~\hyperref[Fig:Literaturesequences]{7(b)}), $T_2$ increases
steadily, indicating suppression of dipolar interactions and
convergence of the average Hamiltonian to zero, similar to the
homogeneous case ($\Delta\nu_i\ll\nu_{ij}$, Supplemental
Fig.~\hyperref[Fig:Literaturesequences]{7(a)}). The interplay
between the opposing effects of the reappearance of the flip-flops
and the convergence of the average Hamiltonian depends strongly on
the magnitude of the quadrupolar inhomogeneity and rf pulse
duration $t_\pi$. However, it is possible to establish a
qualitative agreement between the experiment and the simulations:
for a wide range of the CHASE-10/20 cycle numbers ($\lesssim200
t_\pi$ in Supplemental
Fig.~\hyperref[Fig:Literaturesequences]{7(b)}), $T_2$ is nearly
constant -- this matches the weak dependence of the experimentally
measured $T_2$ on the number of cycles as observed in Figs~2(c,d)
of the main text.

We now examine the effect of the finite `soft' pulses ($t_\pi>0$)
under strong inhomogeneous broadening conditions
($\Delta\nu_i\gg\nu_{ij}$, solid symbols in Supplemental
Figs~\hyperref[Fig:Literaturesequences]{7(b),(d)}). It follows
from Supplemental Fig.~\hyperref[Fig:Literaturesequences]{7(d)}
that the loss of transverse spin polarisation during the control
rf pulses (observed as decrease in the initial echo amplitude
$\langle I_\mathrm{z}(\tau_\mathrm{evol}=0)\rangle$) is most
pronounced for the MS sequences -- the spin coherence can be
maintained well above $\langle
I_\mathrm{z}(\tau_\mathrm{evol}=0)\rangle \sim 0.7$ only for one
cycle of MS-7. One cycle of MKL-68 with a total gate time of $36
t_\pi$ can preserve the echo amplitude above the 70\% threshold
but the resulting coherence time $T_2<2$~ms is shorter than for
Hahn echo. By contrast, the CHASE sequences demonstrate the best
performance in terms of both preserving the echo amplitude
$\langle I_\mathrm{z}(\tau_\mathrm{evol}=0)\rangle$ under long rf
pulse trains and enhancing the coherence time $T_2$. While
CHASE-20 can maintain $\langle
I_\mathrm{z}(\tau_\mathrm{evol}=0)\rangle>0.7$ for gate times
$>100 t_\pi$, the coherence time $T_2$ decreases abruptly above
$24 t_\pi$ in case of the `Y' initialisation pulse. A robust
performance in terms of `freezing' of the spin bath fluctuation
using finite pulses is obtained for either CHASE-10/20 or CHASE-34
for the total rf pulse gate times up to $20-24 t_\pi$ with
CHASE-34 producing a longer coherence time $T_2$.

\subsubsection{\label{sec:sn4-5-3} Analysis and discussion.}

In various applications of magnetic resonance it is a common aim
to seek for an optimal shape of the rf control field that produces
the desired spin manipulation\cite{khaneja2001s,souza2011s}. It
is thus useful to compare the different pulse sequences discussed
here by introducing a quantity that characterises their
efficiency. To this end we take the ratio of the coherence time
$T_2$ during free evolution and the duration of the rf control
pulses required to achieve such $T_2$ -- in other words, the pulse
sequence is more efficient if it yields increased $T_2$ at reduced
overhead of spin manipulation via the rf control pulses. The
dashed lines in Supplemental
Figs~\hyperref[Fig:Literaturesequences]{7(a),(b)} show constant
efficiency levels (given by linear functions with different
slopes). It follows from Supplemental
Fig.~\hyperref[Fig:Literaturesequences]{7(a)} that in case of zero
inhomogeneous resonance broadening ($\Delta\nu_i\ll\nu_{ij}$) the
efficiency is nearly invariant, gradually decreasing with the
growing number of sequence cycle repeats. In case of large
inhomogeneity ($\Delta\nu_i\gg\nu_{ij}$) the increase in the
number of sequence cycle repeates (total gate time) leads to
reduction in efficiency due to the re-appearance of the dipolar
flip-flops discussed above. Supplemental
Fig.~\hyperref[Fig:Literaturesequences]{7(b)} shows that the best
efficiency is achieved for one cycle of either MS-7 or CHASE-10,
while for one cycle of CHASE-34 the coherence time $T_2$ can be
extended only with some loss in efficiency. These results indicate
that when the dipolar-coupled spin bath is inhomogeneously
broadened (Supplemental
Fig.~\hyperref[Fig:Literaturesequences]{7(b)}) its coherence can
be extended efficiently only by introducing complex pulse
sequences that cancel higher order terms of the averaged spin
Hamiltonian -- this is different from the case of zero
inhomogeneous broadening (Supplemental
Fig.~\hyperref[Fig:Literaturesequences]{7(a)}) where cycles of the
basic sequence repeated multiple times efficiently enhance the
spin bath coherence.

Further improvements in simultaneous
suppression of spectral broadening and dipolar couplings in
nuclear spin baths may benefit from techniques beyond AHT. One
example of such a technique are composite pulses. We have conducted
preliminary numerical simulations with modified CHASE sequences,
where each pulse is replaced by a composite broadband BB1 pulse
\cite{wimperis1994s}. However, these pulses give no improvement
and in fact result in a slight reduction of the nuclear spin
coherence times $T_2$, while requiring significantly longer gate
times (and hence reduced efficiency). Alternative approaches may
involve more sophisticated tools, including numerical optimization
algorithms \cite{khaneja2005s}.

To summarise the results of these numerical simulations, we find
that optimal control of the spin bath coherence is achieved using
one cycle of the CHASE-10, CHASE-20, or CHASE-34 sequences as they
(i) extend the spin bath coherence time $T_2$ both under small and
large inhomogeneous resonance broadening, (ii) show robust
preservation of the spin bath magnetisation even under non-ideal
finite duration (`soft') control pulses, (iii) effectively
`freeze' nuclear spin fluctuations regardless of their direction
in the plane perpendicular to the external magnetic field. The
simulations for CHASE-34 predict significant improvement of the
coherence compared to CHASE-10/20 when applied to an
inhomogeneously broadened system -- this is confirmed in
experiment on nuclear spins in self-assembled quantum dots,
although in the experiment the loss of spin bath magnetisation is
found to be larger than expected from simulations, most likely due
the accumulation of amplitude and/or phase errors of the real rf
pulses. Overall, the CHASE sequences developed here provide a well
balanced performance and can be used to control spin bath
fluctuations both in systems with large inhomogeneous resonance
broadening (e.g.\ quantum dots) and systems with small broadening
(e.g.\ defect spins in diamond) where good tolerance to non-ideal
finite pulses is required.


\end{document}